\documentclass[aoas]{imsart}

\RequirePackage{amsthm,amsmath,amsfonts,amssymb}
\RequirePackage[authoryear]{natbib}
\RequirePackage[colorlinks,citecolor=blue,urlcolor=blue]{hyperref}
\RequirePackage{graphicx}
\RequirePackage{bm}
\RequirePackage[linesnumbered,ruled,vlined]{algorithm2e}

\startlocaldefs

\endlocaldefs

\begin{document}

\begin{frontmatter}
\title{Inference for extreme earthquake magnitudes accounting for a time-varying measurement process}
\runtitle{Inference for extreme earthquake magnitudes}

\begin{aug}
\author[A]{\fnms{Zak} \snm{Varty}\ead[label=e1]{z.varty@lancaster.ac.uk}},
\author[A]{\fnms{Jonathan A.} \snm{Tawn}\ead[label=e2,mark]{j.tawn@lancaster.ac.uk}}
\author[A]{\fnms{Peter M.} \snm{Atkinson}\ead[label=e3,mark]{pma@lancaster.ac.uk}}
\and
\author[B]{\fnms{Stijn} \snm{Bierman}\ead[label=e4,mark]{stijn.bierman@shell.com}}
\address[A]{Lancaster University,
\printead{e1,e2,e3}}

\address[B]{Shell Global Solutions Netherlands,
\printead{e4}}
\end{aug}

\begin{abstract}
Investment in measuring a process more completely or accurately is only useful if these improvements can be utilised during modelling and inference. We consider how improvements to data quality over time can be incorporated when selecting a modelling threshold and in the subsequent inference of an extreme value analysis. Motivated by earthquake catalogues, we consider variable data quality in the form of rounded and incompletely observed data. We develop an approach to select a time-varying modelling threshold that makes best use of the available data, accounting for uncertainty in the magnitude model and for the rounding of observations. We show the benefits of the proposed approach on simulated data and apply the method to a catalogue of earthquakes induced by gas extraction in the Netherlands. This more than doubles the usable catalogue size and greatly increases the precision of high magnitude quantile estimates. This has important consequences for the design and cost of earthquake defences. For the first time, we find compelling data-driven evidence against the applicability of the Gutenberg-Richer law to these earthquakes. Furthermore, our approach to automated threshold selection appears to have much potential for generic applications of extreme value methods.
\end{abstract}

\begin{keyword}
\kwd{earthquake}
\kwd{extreme value methods}
\kwd{magnitude of completion}
\kwd{generalised Pareto distribution}
\kwd{threshold selection}
\kwd{partial censoring}
\end{keyword}

\end{frontmatter}


\section{Introduction}
\subsection{Aims and motivation}
The observational nature of environmental data can lead to challenges during statistical modelling and inference. In particular, improved measurement of an environmental process within a dataset should be acknowledged as part of any inference. Failing to do so leads to biased inference, while including data based only on the initial quality of measurements is overly conservative, leads to inefficient inference, and makes financial investment into the measurement process redundant. We consider how to include changing data quality in an extreme value analysis where low data quality is present as the partial censoring of rounded data. Here and throughout, censored data are values that are missing-not-at-random \citep{little2019statistical}. This paper is motivated by the modelling of earthquake catalogues, but results in a method that is applicable more widely where these data features are present. This new threshold selection method should also be of value in more general extreme value analyses. 
\subsection{Earthquake data}
 Earthquakes are recorded if their locations and magnitudes can be inferred from ground vibrations at sensor locations; this requires an earthquake to be detected by multiple sensors. An earthquake is detected or missed depending on its magnitude and location relative to the sensor network. A low sensitivity network of sensors therefore leads to the partial or complete censoring of small magnitude seismic events. As the network is extended or upgraded over time the censoring of small events is reduced. It is usual in earthquake catalogues for magnitudes to be reported to one decimal place; this data feature is often overlooked during statistical analyses \citep{marzocchi2019how}. Using these rounded, incomplete observations we seek to understand the tail behaviour of the magnitude distribution.  

 Since 1991 the Groningen region of the Netherlands has experienced induced earthquakes. These seismic events are caused by gas extraction and have relatively small magnitudes compared to tectonic events. However, they also occur at much shallower depths than their tectonic equivalents. This means that for equal magnitudes they pose a greater hazard than their tectonic counterparts because their impact is spread over a smaller spatial extent. These small earthquakes are therefore both hazardous and difficult to detect. This has led to continued investment in the geophone network around the Groningen gas field to increase detection of small earthquakes and to better understand earthquake activity in the region. Estimating high quantiles of the magnitude distribution, and quantifying their uncertainty, is instrumental to appropriate design and improvement of buildings to withstand these earthquakes.
 
\subsection{Magnitude of completion}
The magnitude of completion $m_c$ is the lowest magnitude above which all earthquakes are certain to be recorded in a given area and time interval. The magnitude of completion therefore depends on the density and sensitivity of the sensor network as well as the local geology. When a sensor network changes substantially over time $t$, the magnitude of completion in that region can be considered as a function of time, denoted $m_c(t)$. The magnitude of completion is not a quantity that can be determined experimentally, it must be inferred from the set of observed event magnitudes. 

Existing methods for statistical estimation of a constant $m_c$ use parametric or non-parametric methods to detect deviations from the assumed monotonicity of the magnitude distribution \citep{mignan2012estimating}. Parametric methods typically assume an exponential magnitude distribution, based on the empirical magnitude-frequency relationship of \citet{gutenberg1956earthquake}. Heuristic techniques are used to detect deviations from this model based on maximum curvature, goodness-of-fit, or parameter stability.

Several methods exist to estimate a spatially varying magnitude of completion \citep{wiemer2000minimum, mignan2011bayesian}. In contrast, little attention has been given to estimating a changing magnitude of completion over time. Where it has been considered, focus has been on temporary increases in $m_c(t)$ due to residual vibrations following large earthquakes \citep{woessner2005assessing, utsu1995centenary}. Long-term changes in $m_c(t)$ have been considered by assuming a constant value within a pre-determined temporal partitioning \citep{hutton2010earthquake} or a locally constant value estimated using a rolling window \citep{mignan2012estimating}. 

\subsection{Extreme value methods}
To specify a model for earthquake magnitudes we adapt a model from extreme value theory. An asymptotic argument justifies the use of the generalised Pareto distribution (GPD) to model the excesses of a continuous random variable over a suitably chosen threshold, under weak assumptions on the distribution of that random variable \citep{pickands1975statistical}. The distribution function of a random variable $Y$ that follows a GPD, given that it is above the threshold $u$, is
\begin{equation} \label{eq:gpd_distn_function}
    F(y ; \sigma, \xi) = 
    \left \{ 
    \begin{array}{ll}
     1 - [1 + \xi (y-u)/\sigma]_+^{-1/\xi}    & \quad \text{ for }\xi \neq 0, y \geq u, \\
     1 - \exp[- (y-u)/\sigma]   &  \quad \text{ for } \xi = 0 , y \geq u;
    \end{array}
    \right.
\end{equation}
where the shape parameter $\xi \in \mathbb{R}$, scale parameter $\sigma > 0$ and $y_+ = \max(0,y)$. The distribution is exponential when $\xi = 0$, heavy-tailed when $\xi > 0$ and decays to a finite upper end point $y^+ = u - \sigma / \xi$ when $\xi < 0$ \citep{davison1990models}. The GPD generalises the Gutenberg-Richter model, in which magnitudes are independent and identically distributed (i.i.d.) exponential random variables, by allowing greater flexibility in the tail behaviour of the distribution. 

Standard extreme value modelling deals with i.i.d. data, observed at regular intervals without rounding or censoring. The standard approach is to select a constant threshold $u$ that is a fixed, high quantile of the empirical distribution. Heuristic methods are used to select an appropriate quantile, see \citet{scarrott2012review} for a review. These methods can be based on the stability of parameter estimates, goodness-of-fit measures, or the mean threshold exceedance size \citep{coles2001introduction}. When interest lies in estimating a particular extreme value property, such as the shape parameter, an alternative strategy is to select the threshold that optimises inference for that property \citep{danielsson2001using}.

Using a constant threshold is inefficient when the data distribution changes over time. This type of change is likely to alter the quantile value above which a GPD is appropriate and cause the GPD parameters to change over time. To avoid this issue, quantiles can be estimated locally as a function of time and a global decision can be made on which quantile to use as a time-varying threshold $u(t)$ \citep{eastoe2009modelling, northrop2011threshold}.

\subsection{Shortcomings of current methods}
Estimating the magnitude of completion and selecting an extreme value threshold are closely linked problems. Both aim to select a value (possibly time-varying) above which a probability model is appropriate. Standard methods from either setting do not meet our modelling needs, for the reasons that follow. 

Methods assuming an exponential magnitude distribution are problematic for two reasons. Firstly, an exponential tail model can lead to bias and false confidence in quantile estimates. \citet{coles2003anticipating} demonstrated in a hydrological context the benefits of using the encompassing generalised Pareto model to properly represent uncertainty in the tail shape. Secondly, the exponential distribution does not account for rounding of the data, resulting in biased parameter estimates \citep{marzocchi2019how,rohrbeck2018extreme}. Failing to acknowledge this rounding can therefore also cause bias in threshold selection.  

 Methods to select a static threshold are also unsuitable for our problem. To obtain precise estimates of the GPD parameters and high quantiles, as much data as possible should be used in the analysis. However, this must be balanced by the need to represent model uncertainty and avoid bias from incorrectly including small magnitude events. This bias has two sources: using either data values for which the extreme value model does not apply or values that are below the magnitude of completion at the time of their occurrence. The optimal choice of time-varying threshold is therefore $v(t) = \max\{m_c(t), u(t)\}$.  Methods for selecting a static modelling threshold $v$ are inefficient when the true threshold varies with time, since the static threshold must satisfy $v \geq \max_t v(t)$ and so excludes viable data from the analysis.
 
 Finally, current approaches to selecting or estimating time-varying thresholds are also unsuitable for our problem; methods for estimating $m_c(t)$ consider only a small portion of the data at once, while the selection of $u(t)$ by a local quantile approach is impeded by the temporal development of the censoring process.
 
\subsection{Contributions and outline}

In this paper we develop an automated method to select a dynamic threshold for rounded GPD data. This is, to our knowledge, the first time that data rounding has been considered during threshold selection. Our proposed threshold selection method uses as much data as possible while guarding against the use of values where a tail model is not appropriate or observations are not complete. This threshold choice leads to more precise estimation of high magnitude quantiles, properly represents their uncertainty, and can also suggest how the magnitude of completion changes over time. The selection method is developed for earthquake data, but the core idea of the method can be applied to extreme value threshold selection more generally. We demonstrate, via simulation, the benefits of including additional, small magnitude events in an extreme value analysis to both parameter recovery and return level estimation. We go on to select dynamic thresholds for partially censored earthquake catalogues and investigate the impact of this threshold when estimating high quantiles of the magnitude distribution.

This paper is structured as follows. Section~\ref{sec:motivating_data} describes the Groningen earthquake catalogue that motivates the proposed methodology, the model for observed magnitudes, and the novel inference for the underlying parameters. Section~\ref{sec:variable_threshold_benefits} demonstrates the benefits of including small magnitude events into an extreme value analysis. Section~\ref{sec:threshold_selection} introduces our proposed method of threshold selection. The method is applied to simulated earthquake catalogues in Section~\ref{sec:simulated_application} and to the Groningen catalogue in Section~\ref{sec:groningen_application}. Concluding remarks are given in Section~\ref{sec:conclusion}.

\section{Motivating data and model formulation} 
\label{sec:motivating_data}
\subsection{Data description}
We study the induced earthquakes in the Groningen region of the Netherlands from January 1$^{\text{st}}$ 1995 to December 31$^{\text{st}}$ 2019. Compared to tectonic earthquakes, these  are close to the surface and can cause damage despite their relatively small magnitudes. This has led the Royal Dutch Meteorological Institute (KNMI) to invest heavily in the earthquake detection infrastructure in the Groningen region. Over time, more and better sensors have been added in the region to increase the detection and reporting of small earthquakes. The resulting earthquake catalogue is publicly available and magnitudes are reported in units of local magnitude ($\text{M}_\text{L}$) to one decimal place \citep{KNMIjson}.

\begin{figure}[htbp]
    \centering
        \includegraphics[width = 0.4\textwidth]{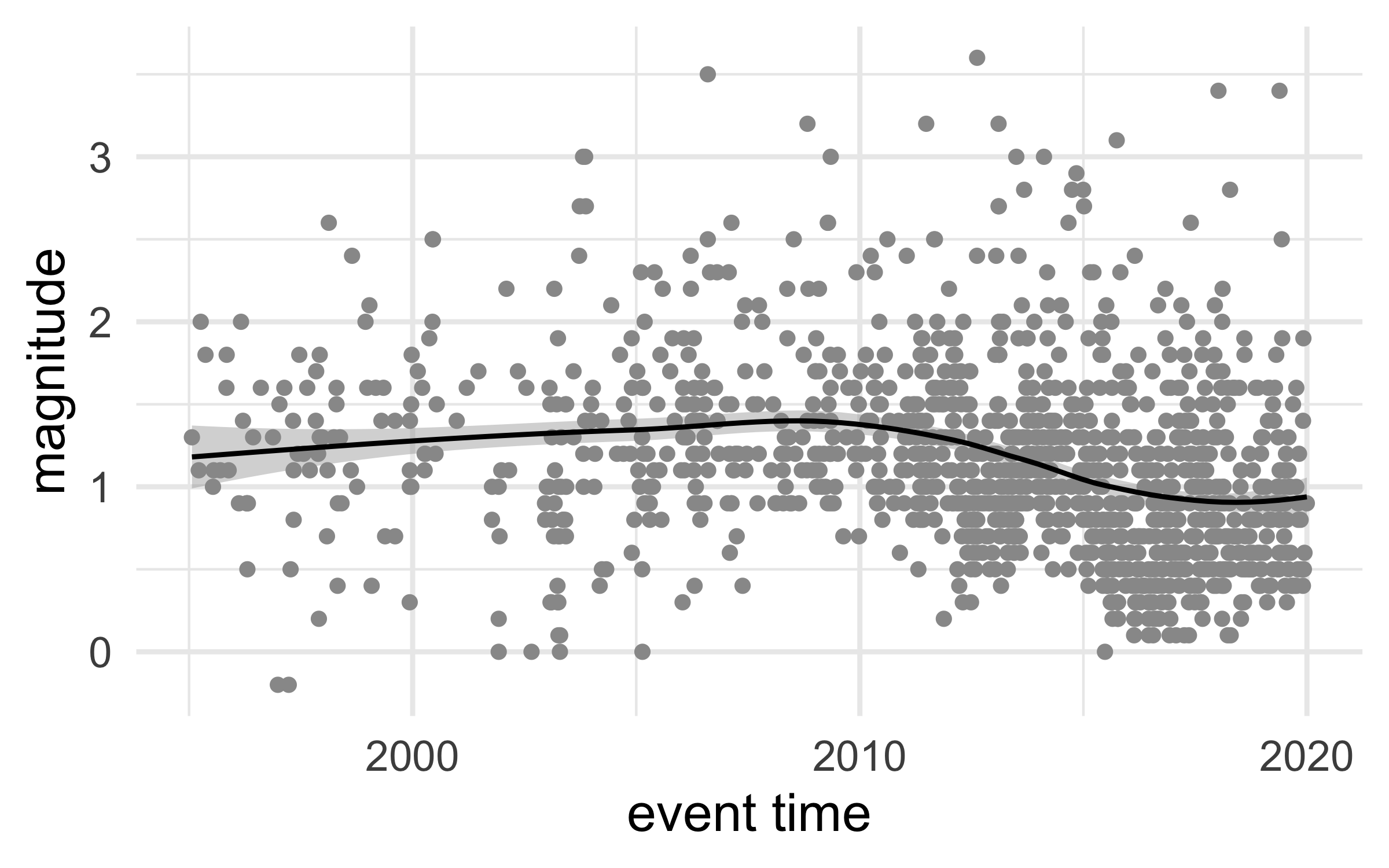}
        \qquad
        \includegraphics[width = 0.4\textwidth]{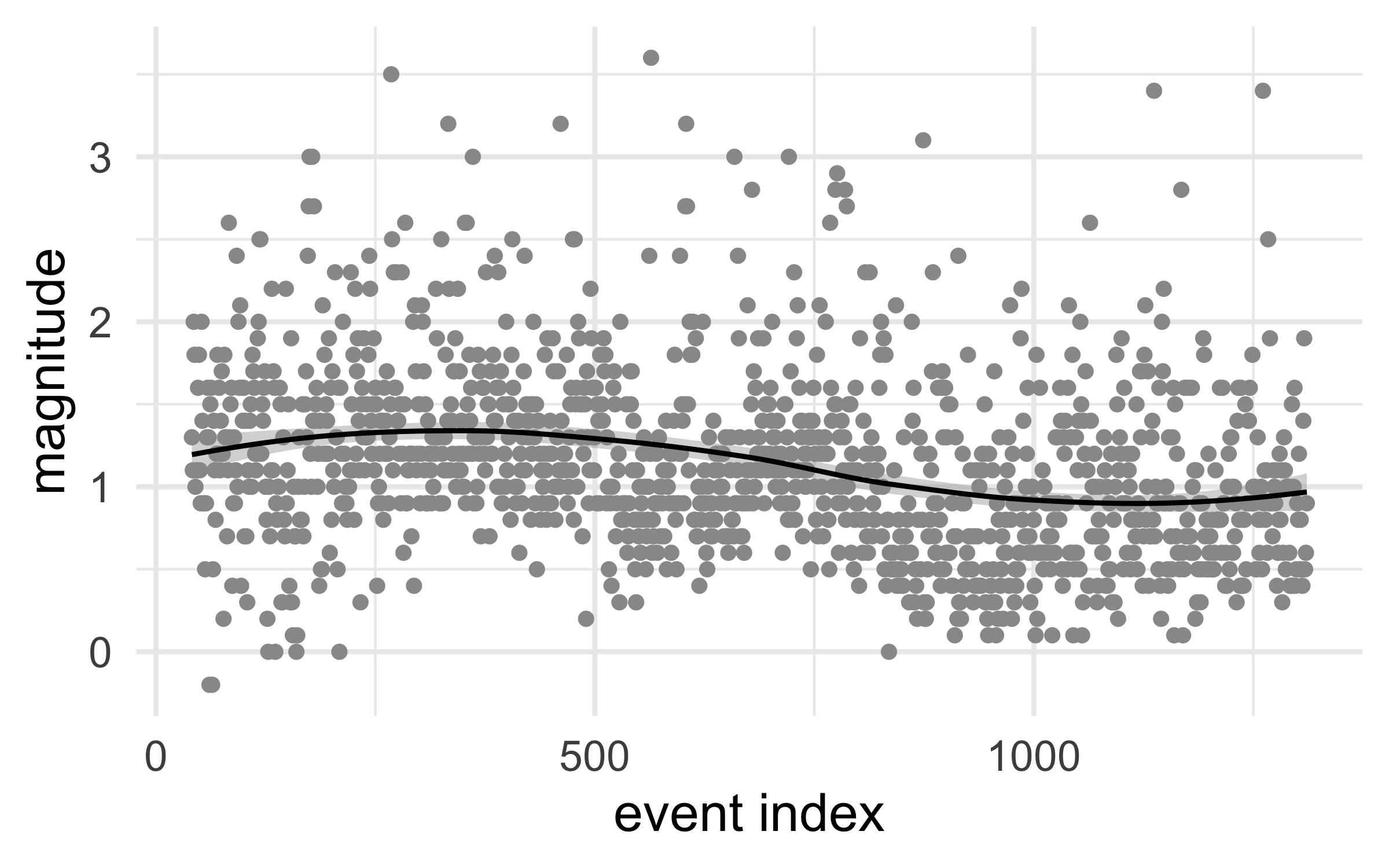}
    \caption{Full Groningen earthquake catalogue, with magnitudes reported in $\text{M}_{\text{L}}$ and smoothed mean estimate; shown using natural- [left] and index-times [right].}
    \label{fig:groningen_catalogue}
\end{figure}

Figure~\ref{fig:groningen_catalogue} shows Groningen earthquake magnitudes against both occurrence time and earthquake index, along with smoothed estimates of their mean using a generalised additive model with cubic-spline basis. Assuming that magnitudes are i.i.d. (which is supported by the exploratory analysis in Section~A of the Supplementary Materials \citep{varty2021inference}) and that departures from this are due to the partial censoring of small magnitude events, the reduction in mean magnitude indicates that fewer small magnitude events were censored at later times. It is unclear whether this change in detection was sudden or gradual. The KNMI report that $m_c(t) \leq 1.5\text{M}_{L}$ for the entire period \citep{dost2012monitoring}. \citet{paleja2016measuring} and \citet{dost2017development} used a fixed temporal partitioning and conclude that for the period 2014-09-24 to 2016-09-27 the magnitude of completion was likely to be below 1.0$\text{M}_\text{L}$. Since sensors have not been removed from the network, this suggests that the magnitude of completion should be less than or equal to this in the period following their analysis (i.e. to 2020 in Figure~\ref{fig:groningen_catalogue}).  

\subsection{Data model and inference} \label{sec:data_model_and_inference}
This section introduces our notation and data model for threshold selection and inference on extreme earthquake magnitudes. We define an earthquake catalogue to be the set of $n$ recorded time-magnitude pairs $\{(t_i, x_i): i = 1 , \dots, n\}$ where the recorded magnitudes $\bm{x} = (x_1, \dots, x_n)$ are given rounded to the nearest $2\delta$ ($\delta >0$) and the event times $\bm{t} = (t_1, \dots, t_n)$ are each within the observation interval $( t_{\text{min}}, t_{\text{max}} )$. The unrounded magnitudes associated with each event are represented by the vector $\bm{y} = (y_1, \dots, y_n)$. An event $(t_i, x_i)$ therefore corresponds to an earthquake of magnitude $y_i \in (x_i - \delta, x_i + \delta]$ that occurred at time $t_i$ and that was not censored.  

Recall from Figure~\ref{fig:groningen_catalogue} that earthquake intensity is not constant over the observation period. To separate exposition of our threshold selection method from estimation of this temporally-varying earthquake rate, we map each event time to its corresponding index. This transforms event times $\bm{t}$ from an irregular sequence on the natural timescale $t$ to a regular sequence $\bm{\tau}$ on the index scale $\tau$, where observed events occur at $\tau = 1, \dots, n$. A modelling threshold $v(\tau)$ is then specified for the transformed observation period $\tau \in (0, \tau_{\text{max}})$ and the threshold values at each event time are given by the vector $\bm{v} = (v(1), \dots, v(n)) = (v_1, \dots, v_n)$. The threshold function $v(\tau)$ and threshold vector $\bm{v}$ will be treated as known until threshold selection is discussed in Section~\ref{sec:threshold_selection}.

The probability $\alpha(\tau, y)$ that an event is detected by the sensor network and included in the earthquake catalogue is an unknown function of its time and magnitude. It is expected that for the Groningen catalogue $\alpha(\tau, y)$ is a non-decreasing function in each of $\tau$ and $y$; larger or later earthquakes are more likely to be detected. We make two assumptions on $\alpha(\tau, y)$: firstly that observation is complete above the modelling threshold, so that $\alpha(\tau, y) = 1$ for $y \geq v(\tau)$; secondly that censoring begins gradually so that for all $\tau$, $\alpha(\tau, y) \approx 1 $ when $y \in [v(\tau) - \delta, v(\tau)]$. This allows rounded magnitudes within $\delta$ of the modelling threshold to be included during inference without constructing a full model for the censoring process. 

In constructing our model for magnitudes exceeding $v(\tau)$, we assume that the unrounded magnitudes $\bm{y}$ may be modelled as i.i.d. GPD random variables $(Y_1, \dots, Y_n)$ with parameters $\bm{\theta} = (\sigma_u, \xi)$ when they exceed a constant, lower threshold of $u < \min_\tau v(\tau) - \delta$. Formally, we assume $Y_i - u | Y_i > u \sim \text{GPD}(\sigma_u, \xi)$. Since events exceeding $v(\tau)$ are never censored, excess magnitudes of $v(\tau)$ may also be modelled using a GPD  but with threshold dependent scale parameters $\sigma_{v_i} = \sigma_u + \xi (v_i - u)$, so that $Y_i - v_i | Y_i > v_i \sim \text{GPD}(\sigma_{v_i}, \xi)$.  

When using this probability model to construct a likelihood function for the GPD parameters, rounded magnitudes $x_i$ should contribute only if the latent value $y_i > v_i$. Events with $x_i > v_i + \delta$, should certainly contribute to the likelihood function and events with $x_i < v_i - \delta$ certainly should not. When $|x_i - v_i| < \delta$ it is uncertain whether $y_i > v_i$ and whether event $i$ should contribute to the likelihood. Each event is therefore weighted in the log-likelihood by $w_i = \Pr(Y_i > v_i | x_i, \bm{\theta})$, the probability it truly exceeds $v(\tau)$. This is equivalent to using the expected likelihood over all possible unrounded magnitude vectors. The resulting log-likelihood function for the the parameters $\bm{\theta} = (\sigma_u, \xi)$ of $F$, the GPD \eqref{eq:gpd_distn_function} is: 
\begin{align} 
    \ell(\bm{\theta} | \bm{x}, \bm{v})
    &= \sum_{i = 1}^{n} w_i \log \Pr(X_i = x_i | Y_i > v_i, \bm{\theta}) \nonumber\\
    & = \sum_{i = 1}^{n} w_i \log \Pr( \max(v_i, x_i - \delta) < Y_i < x_i + \delta | \bm{\theta}) \label{eqn:rounded_gpd_likelihood}\\
    & = \sum_{i = 1}^{n} 
  w_i
    \log\left[ 
         F(x_i + \delta - v_i; \sigma_{v_i}, \xi) - F(\max(v_i, x_i - \delta) - v_i; \sigma_{v_i}, \xi)
        \right], \nonumber
\end{align}
where 
\begin{align}
    w_i &=  \frac{\Pr( \max(v_i,x_i - \delta) < Y_i < x_i + \delta | \bm{\theta})}{\Pr( x_i - \delta < Y_i < x_i + \delta | \bm{\theta})}  \nonumber\\
    \vspace{2em}
    &= \frac{F(x_i + \delta - u; \sigma_u, \xi) - F(\max(v_i, x_i - \delta) - u; \sigma_u, \xi)}
         {F(x_i + \delta - u; \sigma_u, \xi) - F(x_i - \delta - u; \sigma_u, \xi)} . \label{eqn:rounded_gpd_weights}
\end{align}

 The maximum likelihood estimate $ \bm{\hat \theta}$ can be found using numerical optimisation of this function. Confidence intervals may be obtained based on asymptotic normality, but this approximation can be poor for the estimated shape parameter $\hat \xi$ and quantile values. To avoid this and to ensure that confidence bounds on $\hat \sigma_u$ are positive, we use a parametric bootstrap approach to describe parameter uncertainty, as described in Section~B of the Supplementary Materials \citep{varty2021inference}. 

\section{Motivating the inclusion of small magnitudes} 
\label{sec:variable_threshold_benefits}
\subsection{Simulation study overview} 
Here we show that using a non-constant threshold to include additional, small magnitude earthquakes in an extreme value analysis can be beneficial to both parameter and quantile estimation. We compare three approaches to inference on 1000 simulated earthquake catalogues that have a known, stepped threshold. Each catalogue is simulated by first generating 1000 latent magnitudes as independent GPD exceedances of $u = 1.05 M_{\text{L}}$ with parameters $\bm{\theta} = (\sigma_u, \xi) = (0.4,0.1)$. Each event $i = 1,\dots, 1000$ is censored if $\tau_i \leq 500$ and $y_i < 1.65  M_{\text{L}}$. The retained magnitudes are then rounded to the nearest $2\delta = 0.1 M_{\text{L}}$, resulting in a catalogue of the form shown in Figure~\ref{fig:sim_cat_structure_and_return_levels} (left). The size of the retained catalogue depends on the simulated magnitudes, and so varies between catalogues.  

A GPD model is fitted to each of the simulated catalogues by maximising the log-likelihood \eqref{eqn:rounded_gpd_likelihood} under each of three approaches. The first, conservative approach to inference uses only exceedances of the flat modelling threshold $v(\tau) = 1.65 M_{\text{L}}$ for $0 \leq \tau \leq 1000$. The second approach uses exceedances of the stepped threshold where $v(\tau) = 1.65 M_{\text{L}}$ for $0 \leq \tau \leq 500$ and $v(\tau) = 1.05 M_{\text{L}}$ for $500 < \tau \leq 1000$. The number of data points used by the stepped approach will be at least as large as by the conservative approach. A third approach, possible in simulation but not practice, is also considered. In this third approach, additional earthquakes are simulated above the conservative level to extend the simulated catalogue until the number of exceedances of $1.65 M_{\text{L}}$ matches the number of events used by the stepped approach. A GPD model is then fitted to the extended set of earthquakes that exceed $1.65 M_{\text{L}}$. 

We compare the three approaches to inference in terms of parameter and quantile estimation. The conclusion of each comparison can differ because of the non-linear relationship between GPD parameters and quantiles, which are also sensitive to small changes in the estimated shape parameter $\xi$. Parameter estimates are compared using their bias and variance over the 1000 simulated catalogues. To be able to compare quantile estimates across modelling thresholds we consider the conditional quantiles above the conservative threshold level, using conditional return levels. The conditional $p$-quantile above some magnitude $c > u$ is the magnitude $y_{p,c}$ that satisfies 
\begin{equation*}
    \Pr(Y \leq y_{p,c} | Y > c) = p.
\end{equation*}
Letting $\zeta_c = \Pr(Y > c | Y >u) = 1 - F(c ; \bm{\theta})$, where $F$ is the distribution function \eqref{eq:gpd_distn_function}, $y_{p,c}$ can be expressed as a function of $\bm{\theta}$:
\begin{equation}
    y_{p,c}(\bm{\theta}) = 
    \left\{
    \begin{array}{ll}
        u + \frac{\sigma_u}{\xi}\left((\zeta_c p)^{-\xi} - 1 \right) &  \text{for } \xi \neq 0, \\
        u + \log(\zeta_c p) & \text{for } \xi = 0. 
    \end{array}
    \right. 
    \label{eqn:conditional_quantile_definition}
\end{equation}
An alternative representation of conditional quantiles, more in-keeping with the extreme value approach, is the $m$-event conditional return level above $c$. This can be found by setting $p = 1 - 1/m$ in equation \eqref{eqn:conditional_quantile_definition} and interpreted as the magnitude exceeded (on average) by one in every $m$ events that exceed $c$. We compare point estimates and confidence intervals of conditional return levels under the three approaches to inference.  

\subsection{Simulation study results}
Figure~E.1 of the Supplementary Materials \citep{varty2021inference} shows the sampling distribution of parameter estimates and an error decomposition under each approach to inference. The stepped threshold is best for parameter estimation, with the smallest bias and variance of the three approaches. The mean squared error of the stepped estimator is $9.6$ times smaller than that of the conservative estimator, mainly due to its increased precision. For comparison, artificially extending the earthquake catalogue gives a reduction factor of only $4.2$. In this example, each small magnitude event added by lowering the threshold is more than twice as valuable to parameter estimation than an additional observation above the conservative level. 

\begin{figure}[htbp]
    \centering{}
        \includegraphics[width = 0.55\textwidth, page= 1]{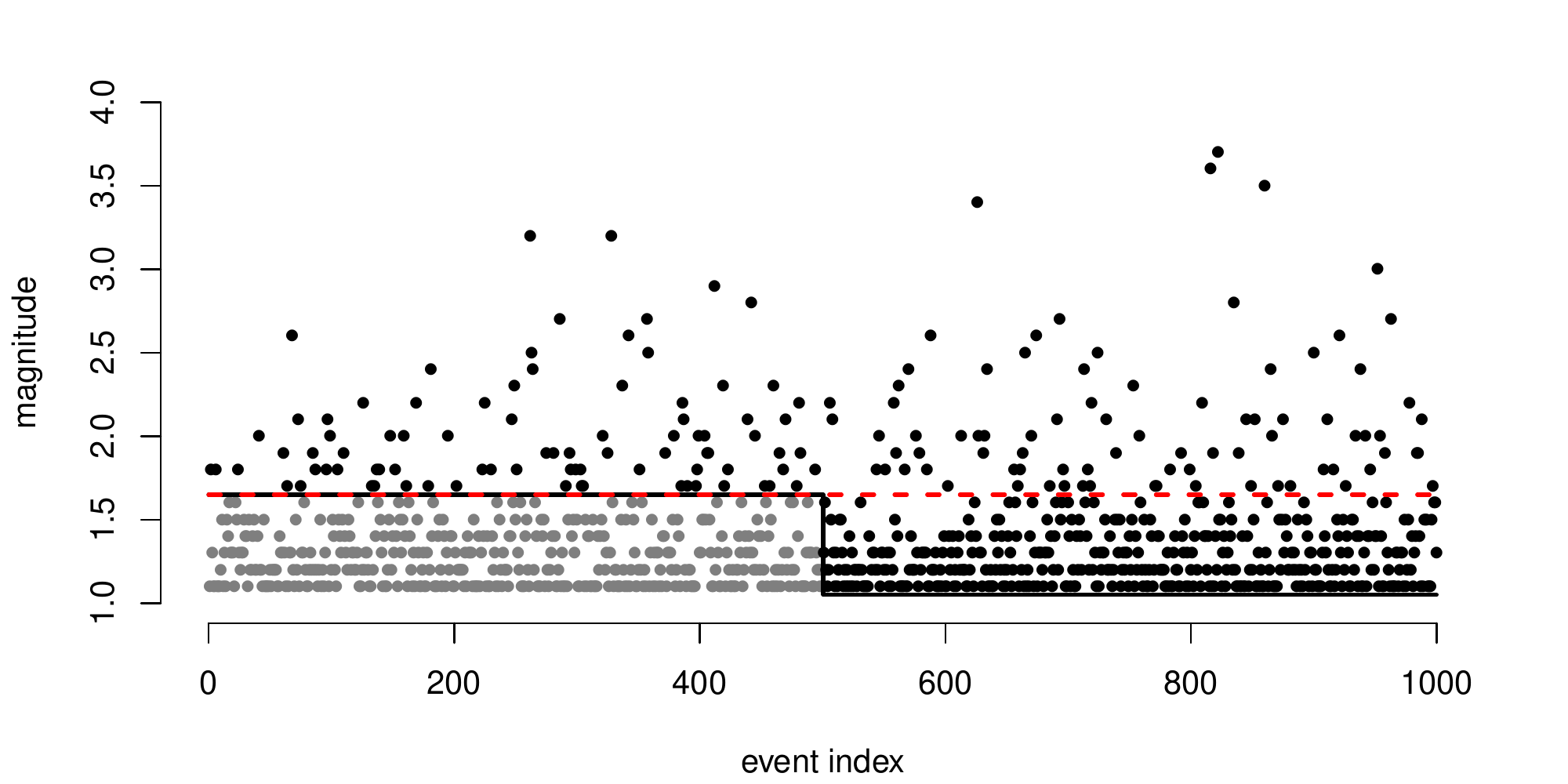}
        \includegraphics[width = 0.4\textwidth, page = 1]{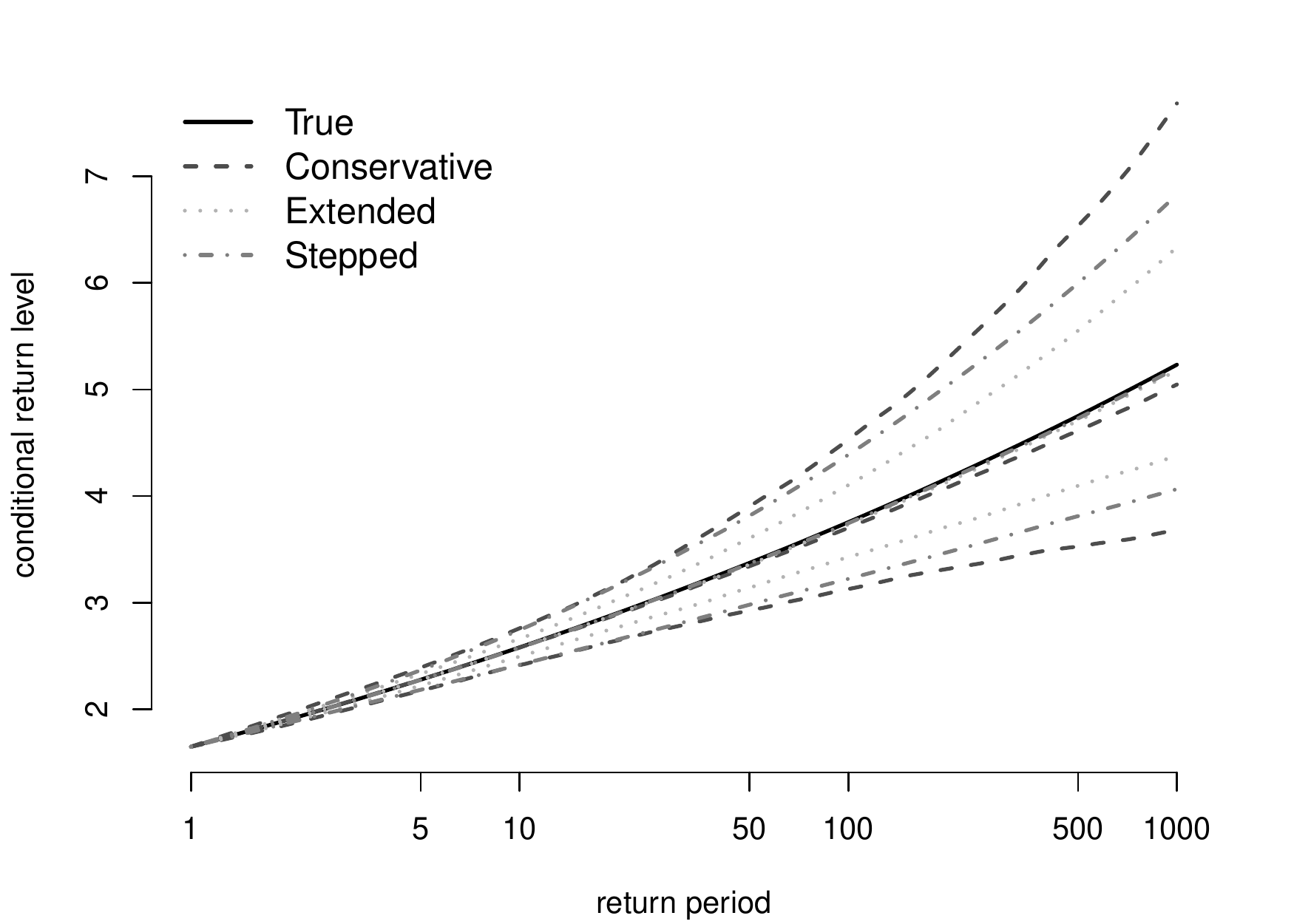} 
        
    \caption{[Left] Simulated catalogue structure: events are censored (grey dots) if in the first 500 and below 1.65$\text{M}_\text{L}$. For this catalogue, the conservative threshold (dashed red line) includes 181 events, while the stepped threshold (solid black line) includes 582 events. [Right] Magnitude conditional return level estimates in $\text{M}_\text{L}$ against return period in number of earthquakes exceeding 1.65$\text{M}_\text{L}$. Point estimates and $95\%$ confidence intervals are given under conservative, extended and stepped approaches to inference, along with the true values.}
    \label{fig:sim_cat_structure_and_return_levels}
\end{figure}

Figure \ref{fig:sim_cat_structure_and_return_levels} (right) shows the conditional return levels for magnitudes above $c = 1.65 \text{M}_{\text{L}}$ under each approach. Point estimates are qualitatively similar in each case, but confidence intervals are narrowed by using the stepped rather than constant threshold. Confidence intervals are further narrowed by artificially extending the observation period. This is because of the additional large values in the extended data, which have a strong influence over the estimated return levels \citep{davison1990models}. 

These results show clearly the benefits for parameter and quantile inference that can be achieved by using a dynamic modelling threshold to include additional small magnitude events in an extreme value analysis. Using a conservative constant threshold leads to wasteful inference and the squandering of these potential gains. 

\section{Threshold selection} \label{sec:threshold_selection}
\subsection{Overview} \label{sec:threshold_selection_overview}
In practice, the true modelling threshold $v(\tau)$ is always unknown. To choose between potential thresholds, we must define what it means for one threshold to be preferred over another. A generalised likelihood ratio test is not appropriate for this comparison because it compares nested models on the same data, rather than comparing the same model on nested data \citep{wadsworth2012likelihood, wadsworth2016exploiting}.

To select a model that is robust to sampling variability, $v(\tau)$ should include as much data as possible in the model and therefore be chosen to be as low as possible. However, selecting $v(\tau) < \max(u(\tau), m_c(\tau))$ for any $0<\tau< \tau_{\text{max}}$ will cause bias in the fitted model, making it incapable of obtaining an asymptotically consistent estimator of the true parameter values. The best choice of $v(\tau)$ is therefore the threshold that includes the most data while maintaining a good agreement between observed threshold exceedances and the fitted GPD.

For i.i.d. continuous valued data, the distributional agreement with a probability model can be assessed graphically by using a PP- or QQ-plot and adding tolerance intervals to show expected behaviour under that model. Alternatively, the distributional fit can be summarised using a metric, such as the Anderson-Darling or Cramer-von Mises distances \citep{laio2004cramer}. Both graphical- and metric-based approaches can be adapted for data $\bm{y}$ that are independent and continuous valued, but which do not have a shared distribution. This is achieved by using the probability integral transform and the fitted distribution to transform the data to have a shared marginal distribution before using methods for i.i.d. data to produce plots or metric values \citep{heffernan2001extreme}. 

We further adapt these methods to handle both rounded data and parameter uncertainty, before showing how they can be used to inform selection of a modelling threshold. In doing so, we transform to standard Exponential margins because this distribution is central within the GPD family and follows the precedent of \citet{heffernan2001extreme}. Alternative marginal distributions could be used; we additionally consider PP-plots, which correspond to the special case of uniform margins.

\subsection{Graphical assessment} \label{sec:threshold_selection_graphical}
The observed magnitudes $\bm{x}$ that exceed $v(\tau)$ do not have a shared marginal distribution when $v(\tau)$ is non-constant and they are not continuous-valued due to their rounding. This presents challenges when trying to create a PP- or QQ-plot for exceedances of the modelling threshold $v(\tau)$. Firstly, constructing these plots using rounded values can lead to many probabilities or quantiles of equal value and the plots being difficult to interpret. The second challenge relates to observed, rounded values close to the modelling threshold, $\{x_i: |x_i - v_i| < \delta, i = 1,\dots, n\}$; it is not known which, or how many, of these events satisfy $y_i \geq v_i$ and so should be included when constructing the plot. 

To overcome these challenges we use simulation to construct Monte Carlo confidence intervals for the sample quantiles (or probabilities) of the unrounded threshold exceedances transformed onto shared exponential margins. The process is described in Section~C of the Supplementary Materials \citep{varty2021inference} and leads to a modified plot with two sets of intervals; tolerance intervals show the expected variability of sample quantiles (or probabilities) under the fitted model while confidence intervals show the uncertainty about the observed sample quantile values. Confidence and tolerance intervals that do not overlap suggest that the distribution of the rounded exceedances is not coherent with the fitted GPD model.

\begin{figure}
    \centering
    \includegraphics[width = 0.25\textwidth, page = 5]{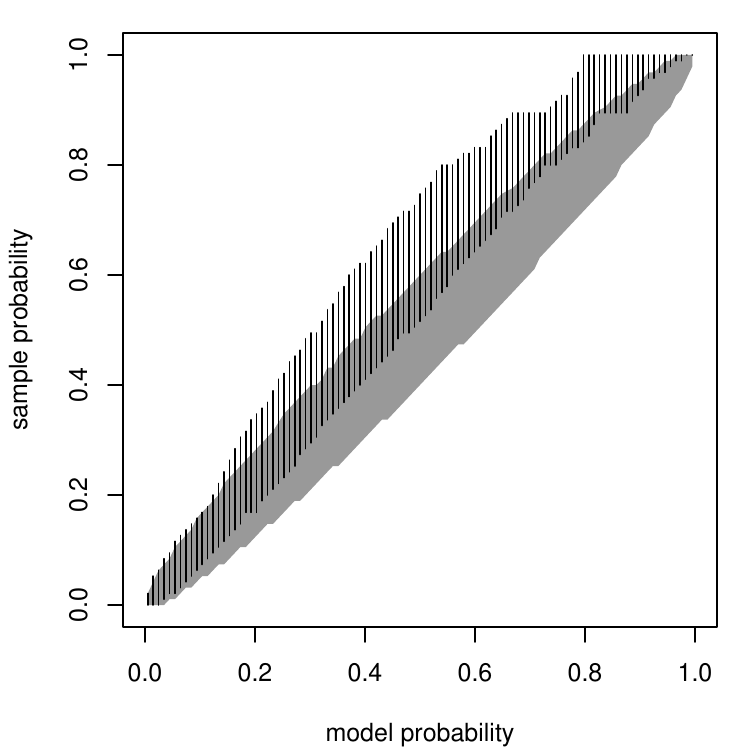}
    \qquad
    \includegraphics[width = 0.25\textwidth, page = 6]{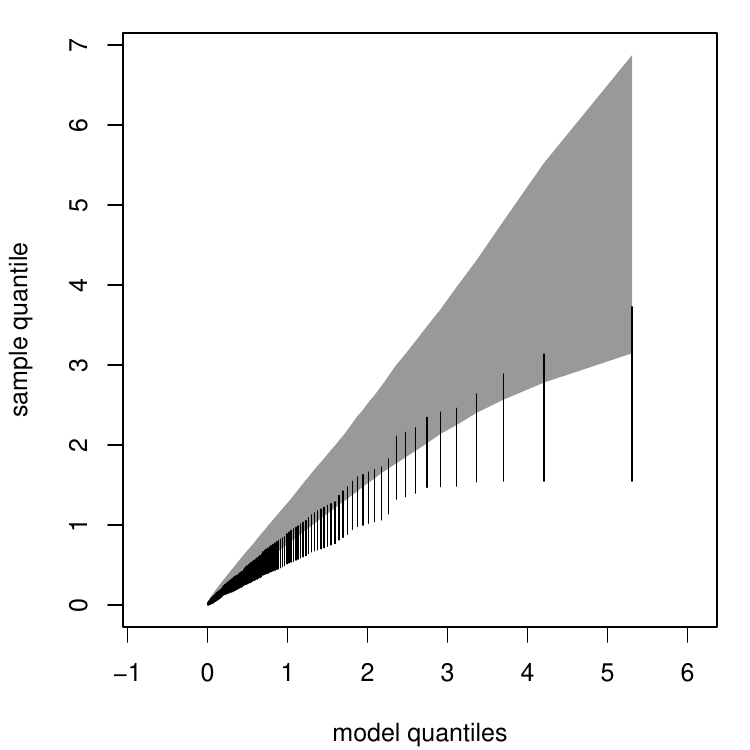} \\ 
     \includegraphics[width = 0.25\textwidth, page = 3]{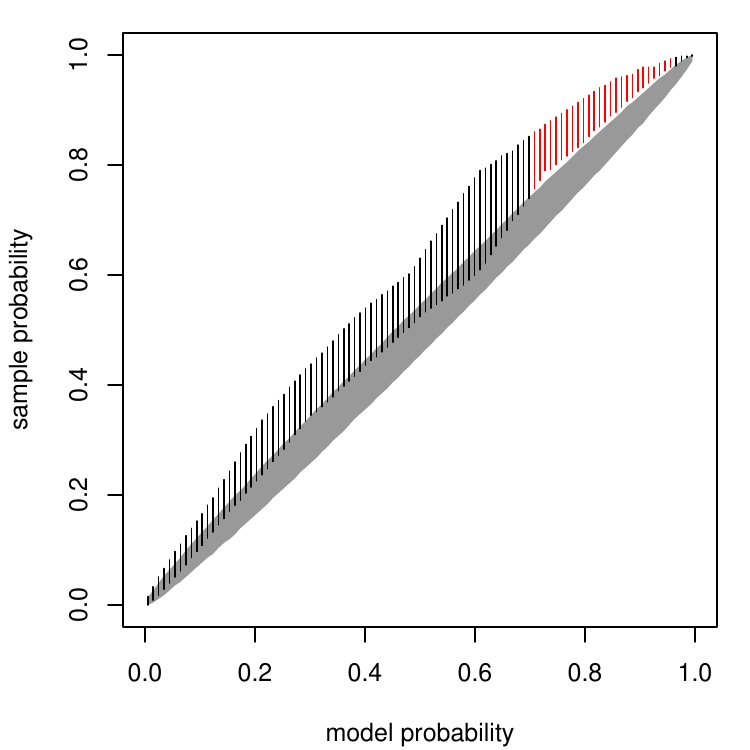}
     \qquad
    \includegraphics[width = 0.25\textwidth, page = 4]{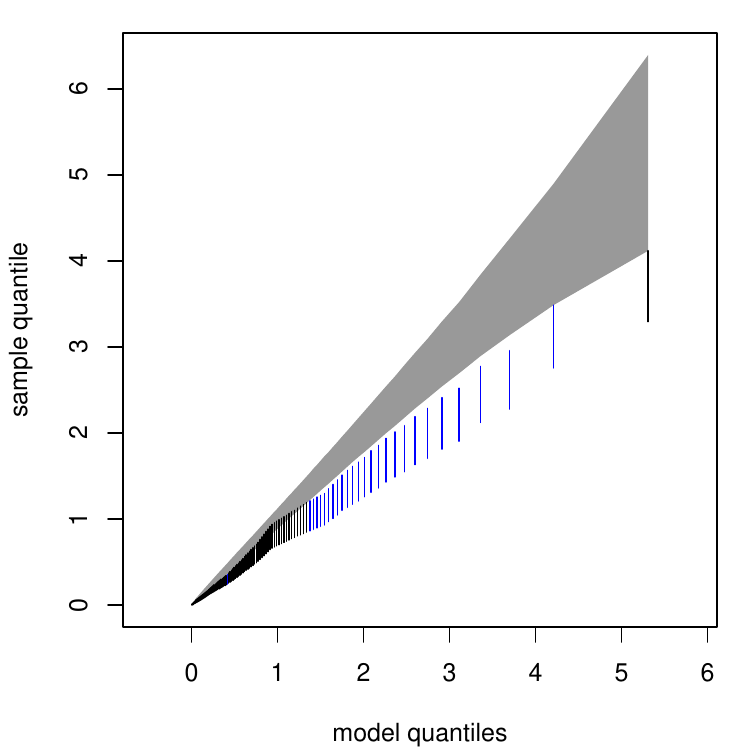}
    \caption{PP-plots [left] and QQ-plots [right] for threshold exceedance sizes shown on Exp(1) margins for constant modelling thresholds $v(\tau) = 1.85\text{M}_{\text{L}}$ [top] and $v(\tau) = 1.15\text{M}_{\text{L}}$ [bottom]. 95$\%$ tolerance intervals are shown as grey regions, while $95\%$ confidence intervals on each probability or quantile are shown as vertical lines. These are coloured red (blue) where the confidence interval is entirely above (below) the tolerance interval.}
    \label{fig:pp_qq}
\end{figure}

Examples of such PP- and QQ-plots are shown in Figure~\ref{fig:pp_qq}. These use the simulated catalogue shown in Figure~\ref{fig:sim_cat_structure_and_return_levels} (left) and constant modelling thresholds of $v(\tau)$ = 1.85$M_L$ and 1.15$M_L$. For this catalogue, exceedances of a flat threshold should be consistent with a GPD model only if that threshold is of $1.65 M_L$ or greater. For the higher threshold $v(\tau)$ = 1.85$M_L$, the confidence intervals on sample probabilities and quantiles overlap with the tolerance intervals, indicating that exceedances of this threshold are consistent with the fitted GPD model. For the lower threshold $v(\tau)$ = 1.15$M_L$ this is not the case, with the large sample quantiles bigger than expected under the fitted model. Notice the shape of the tolerance intervals in Figure~\ref{fig:pp_qq}; the largest deviations from the line $y=x$ are expected at central probabilities in the PP-plots and at the largest quantiles of the QQ-plots. This is important in Section~\ref{sec:threshold_selection_metrical} where we propose metrics to summarise these plots.

\subsection{Metric-based assessment} \label{sec:threshold_selection_metrical}
Using a metric rather than a graphic to assess the distributional coherence of modelled and observed threshold exceedances facilitates the comparison of many thresholds. We therefore aim to summarise the PP- and QQ-plots using metrics that reward accurate estimation of the magnitude distribution function. An unbiased estimate results in a plot that covers the line $y=x$, while a precise estimate results in plots that are stable between sampled values for the mle $\hat{\bm{\theta}}$ and unrounded data $\bm{y}$. Our approach to creating a metric that summarises these plots is novel in its design, which rewards large sample sizes through their effect to increase the precision of the distribution estimate. 

 We propose four metrics to summarise deviation from the line $y=x$ in PP- and QQ-plots using the mean absolute distance and mean squared distance in what follows. The calculation of these metrics is described below for a single sampled vector of threshold exceedances on exponential margins $\tilde{\bm{z}}$. Let $d_0$ be the realised metric value for an arbitrary dataset using one of the four methods, and $d = \mathbb{E}_{\bm{Y},\bm{\hat \theta}|\bm{x},\bm{v}}(d_0)$ be the expected value of $d_0$ over the joint distribution of ${\bm{Y}, \bm{\hat \theta}}| \bm{x}, \bm{v}$, thus accounting for the rounding and parameter uncertainties that are represented by the confidence intervals of Figure \ref{fig:pp_qq}. We select a modelling threshold by minimising $d$ and investigate which choice of $d_0$ is best. Here the expected values of the metrics are calculated by a Monte Carlo approximation.
 
Smaller values of each metric are preferable, with large values caused by the quantiles of the fitted model being either highly uncertain or incoherent with the observed data. Minimising these metrics provides a new approach to threshold selection, which rewards thresholds that give low sampling variability and small bias in the resulting estimator. The remainder of this section covers the calculation of these metrics, while Section~\ref{sec:simulated_application} explores their relative performance on simulated data.
 
In the following, $\tilde{\bm{z}}^{(i)}$ is the $i^{\text{th}}$ parametric-bootstrapped vector of threshold exceedances transformed onto exponential margins for independent, replicated samples $i = 1,\dots, k$. An algorithm to obtain these is given in Section~C of the Supplementary Materials \citep{varty2021inference}. Also let $H_{(i)}(y): \mathbb{R}^+ \rightarrow [0,1]$  and $Q_{(i)}(p) : [0,1] \rightarrow \mathbb{R}^+$, respectively, be the empirical distribution function and the sample quantile function of $\tilde{\bm{z}}^{(i)}$ for $i = 1,\dots, k$. The sample quantile functions are defined as linear interpolations of the points
$\left\{\left(\frac{j-1}{\tilde{n}^{(i)}-1}, \tilde{z}^{(i)}_{(j)}\right): j = 1,\dots,\tilde{n}^{(i)}\right\}$, 
where $\tilde{n}^{(i)}$ is the length of $\tilde{\bm{z}}^{(i)}$ and $\tilde{z}^{(i)}_{(j)}$ is the $j^{\text{th}}$ order statistic of $\tilde{\bm{z}}^{(i)}$.
 
The quantile based distance metrics $d_{(i)}(q,1)$ and $d_{(i)}(q,2)$ summarise the expected deviation in the QQ-plot of $\tilde{\bm{z}}^{(i)}$ from the line $y=x$ at a set of $m \in \mathbb{N}^+$ equally spaced evaluation probabilities $\{p_j = j / (m+1): j = 1,\dots,m\}$. The two metrics respectively give the mean absolute distance and mean squared distance between model and sample quantiles over the set of evaluation probabilities. They are given by
\begin{align*}
    d_{(i)}(q,1) = \frac{1}{m} \sum_{j=1}^{m} |-\log(1-p_j) - Q_{(i)}(p_j)|
\end{align*}
and 
\begin{align*}
    d_{(i)}(q,2) = \frac{1}{m} \sum_{j=1}^{m} (-\log(1-p_j) - Q_{(i)}(p_j))^2.
\end{align*}
%

 In a PP-plot the variance of deviations from the line $y=x$ is greatest when $p_j = 0.5$ and shrinks to 0 as $p_j$ approaches 0 or 1. In the PP-based metrics we therefore weight the sum of the deviations to account for large discrepancies being less surprising for central probabilities. The metrics $d_{(i)}(p,1)$ and $d_{(i)}(p,2)$ are therefore calculated using, respectively, the weighted-absolute and weighted-squared errors: 
\begin{equation*}
    d_{(i)}(p,1) = \frac{1}{m} \sum_{j=1}^{m} 
    \left[ \left(\frac{p_j(1-p_j)}{\sqrt{n^{(i)}}}\right)^{-1/2}
    \left|p_j - H_{(i)}(-\log(1-p_j))\right| \right]
\end{equation*}
and
\begin{equation*}
    d_{(i)}(p,2) = \frac{1}{m} \sum_{j=1}^{m} 
    \left[ \left(\frac{p_j(1-p_j)}{\sqrt{n^{(i)}}}\right)^{-1/2}
    \left(p_j - H_{(i)}(-\log(1-p_j))\right)^2 \right].
\end{equation*}
These deviations are again measured at equally spaced evaluation probabilities $p_1,\dots,p_m$. In the quantile-based metrics the weighting is handled implicitly by choosing equally spaced evaluation probabilities, which gives dense evaluation where discrepancies from $y=x$ are expected to be small and sparse evaluation where they are expected to be large. In this way, the weights reflect the width of the tolerance intervals in Figure~\ref{fig:pp_qq}.

Uncertainties in the estimated GPD parameters, the size of the exceedance set and the values of the unrounded exceedances should all be accounted for when using a metric to select a modelling threshold. This can be achieved by calculating the distance metrics for each of $k$ realisations of the vector $\tilde{\bm{z}}$, where each uses one of $k$ bootstrap parameter estimates of $\bm{\hat \theta}$. The expected metric values over these realisations are denoted by $d(a,b)$, where $a \in \{p,q\}$ and $b \in \{1,2\}$. The expected distance metric $d(q,1)$ is defined as: 
\begin{equation*}
 d(q,1) = \frac{1}{k}\sum_{i =1}^{k} d_{(i)}(q,1),
\end{equation*}
with the other expected distance metrics defined similarly. 
%
%

\subsection{Minimisation procedure} \label{sec:minimisation_procedure} 
To select the most appropriate threshold the threshold parameters which minimise the selected metric $d$ must be found. Standard, gradient-based optimisation procedures are not well suited to this task because the censoring mechanism can cause multiple local minima and the Monte Carlo evaluation leads to local roughness over parameter values. When using a simple parametric form for the threshold, such as a constant or stepped threshold (where the change location is known), a simple grid search can be used to overcome these issues and find the threshold parameters that minimise the metrics. For more complex threshold forms, with a higher dimensional parameter space to optimise over, a grid search becomes prohibitively expensive. 

 To find the threshold parameter set for more complicated thresholds we explore the threshold parameter space in a more principled manner. To do this we use Bayesian optimisation \citep{snoek2012practical} as implemented in the R package \texttt{ParBayesianOptimization} \citep{wilson2020ParBayesianOptimizaion}. The optimisation procedure begins by evaluating $d$ at a small initial collection of randomly chosen parameter vectors within a bounded search space. Based on the resulting metric values, future evaluation points are selected sequentially as the parameter vector with the greatest expected reduction in $d$ as compared to the current best value. This search method balances evaluations between parts of the parameter space where the metric is known to have low values and parts where it is most uncertain.
 
 Bayesian optimisation is a heuristic search method but has been shown in other applications to find good parameter combinations using a relatively small number of function evaluations \citep{shahriari2015taking}. To establish its suitability in our setting we compared Bayesian optimisation to a grid search for two sub-problems; catalogues with a flat threshold and catalogues with a stepped threshold with known change location. In both cases Bayesian optimisation performed favourably compared to grid search, selecting thresholds close to the true value at a lower computational cost. We do not claim that Bayesian optimisation is the best method for optimising the proposed metrics over threshold parameters, only that it appears to be an efficient method of finding good thresholds. 
 
\section{Threshold selection on simulated catalogues} 
\label{sec:simulated_application}
\subsection{Simulation study overview}  \label{sec:simulation_study_overview}
We consider the performance of the proposed threshold selection metrics on a collection of simulated data sets with either constant or stepped threshold forms. This simulation study illustrates the effectiveness of our method and establishes which of the distance metrics proposed in Section~\ref{sec:threshold_selection} is best. 

We attempt to select the most appropriate threshold from a set of candidate thresholds. Two censoring types (hard and phased) are considered for magnitudes that are below the modelling threshold. For hard censoring, all simulated continuous magnitudes below the modelling threshold are undetected. In phased censoring the detection probability of each event, $\alpha(y_i, v_i) = \exp(-\lambda [v_i - y_i]_+)$, decreases as the simulated continuous magnitude falls further below the threshold, as controlled by the parameter $\lambda > 0$. The particular choices of exponential decay and the value of $\lambda$ are arbitrary but were chosen to reflect, in a broad sense, the censoring observed in the Groningen earthquake catalogue. Note that either of these censoring types can result in some rounded magnitudes that are below the threshold even though their simulated continuous values are above the threshold.

\subsection{Constant threshold, hard censoring} \label{sec:constant_threshold_hard_censoring}
\begin{figure}
    \centering
    \includegraphics[width = 0.35\textwidth, page = 2]{figures/flat_threshold_sim/qq_metrics.pdf}
    \includegraphics[width = 0.35\textwidth, page = 1]{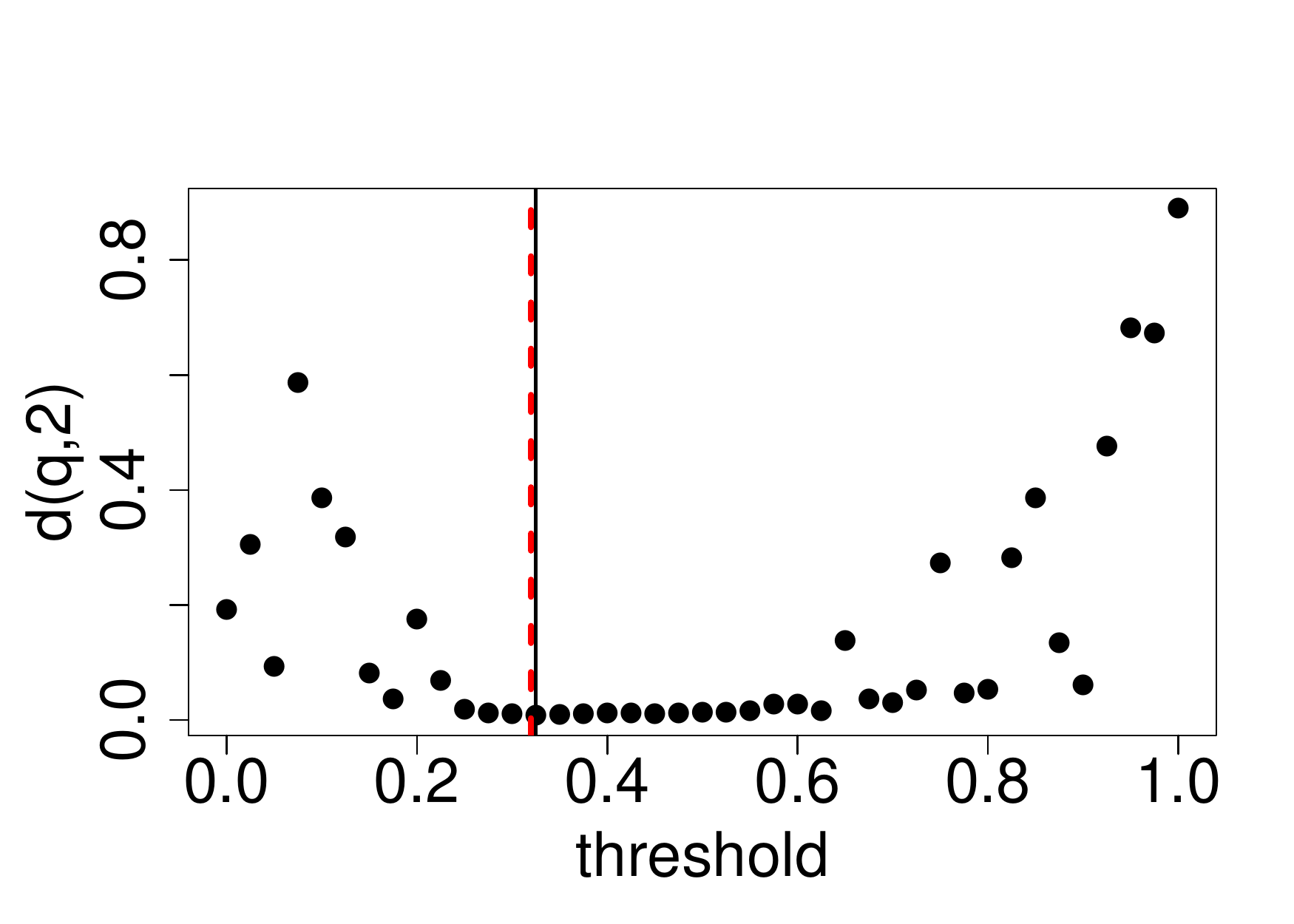} 
    \\
    \includegraphics[width = 0.35\textwidth, page = 2]{figures/flat_threshold_sim/pp_metrics.pdf}
    \includegraphics[width = 0.35\textwidth, page = 1]{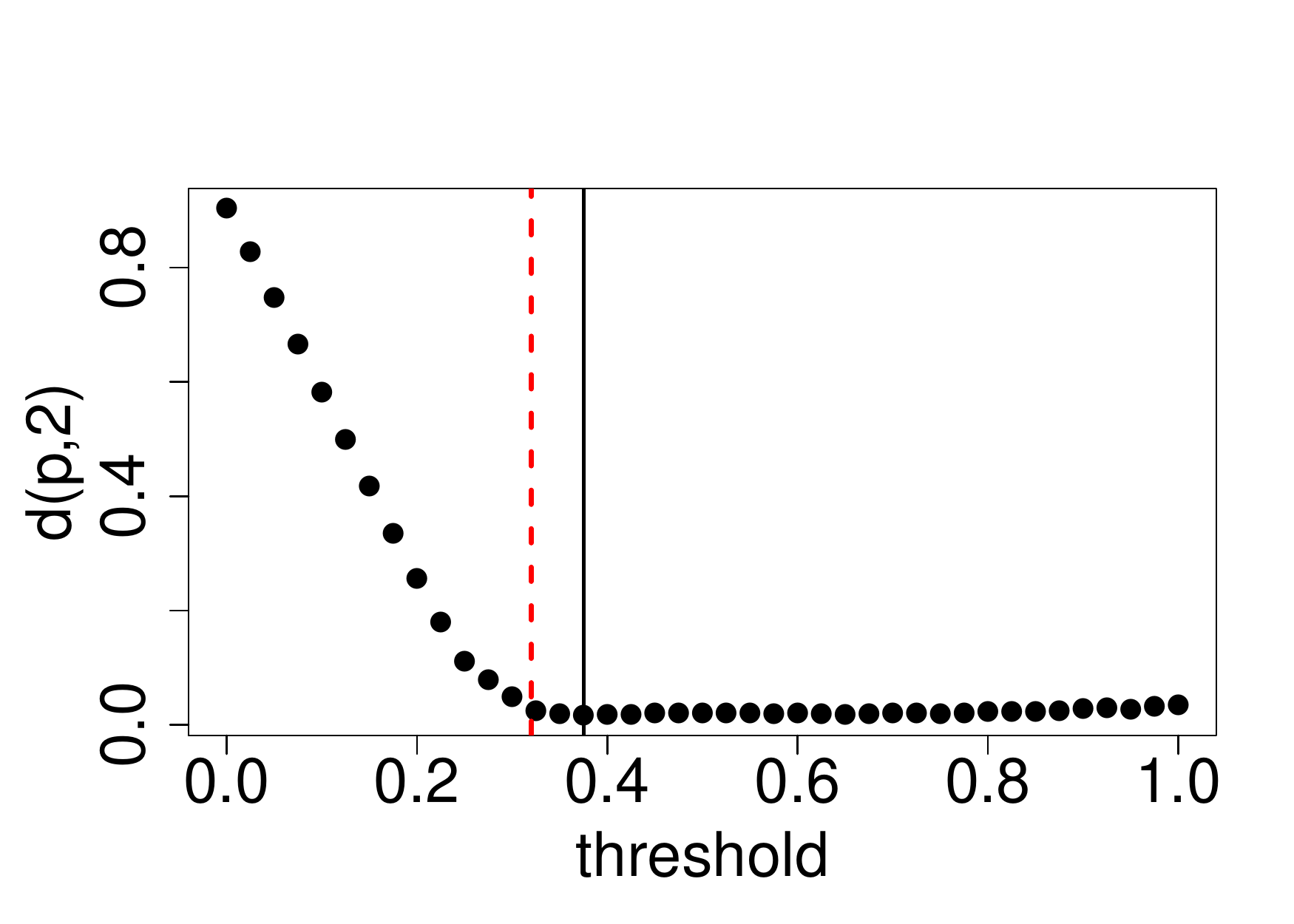} 
    \caption{Flat threshold selection on a simulated catalogue. Top row: expected mean absolute [left] and expected mean squared [right] QQ-distances against threshold value. Bottom row: expected PP-distance metrics based on absolute [left] and squared [right] errors against threshold. Selected and true thresholds are indicated by solid black and dashed red lines.}
    \label{fig:flat_threhsold_sim}
\end{figure}

We first use the four proposed metrics to select a constant threshold for 1500 simulated i.i.d. GPD exceedances of the constant threshold $v(\tau) = 0.32 \text{M}_{\text{L}}$, hard-censored below $v(\tau)$ and rounded to the nearest 0.1$\text{M}_{\text{L}}$.  We first consider the metrics for a single dataset. Expected metric values are calculated at the 41 equally spaced, constant candidate thresholds shown in Figure~\ref{fig:flat_threhsold_sim}. The candidate threshold selected by minimising $d(q,1)$ is the closest threshold on the grid to the true value. This metric also appears to provide the most clearly defined minimum, indicating that it penalises both thresholds that are too low and too high. All four metrics show clear increases in metric value for candidate thresholds that are too low, but not when the candidate threshold is too high. The probability-based metrics do not increase greatly when the candidate threshold is too high, and so fail to adequately reward the inclusion of valid events with smaller magnitudes. This is presumably because they do not sufficiently penalise the increased uncertainty in the estimated parameters when using a higher threshold. 

When selecting a constant threshold, the standard approach is to exploit the well-established property that the GPD shape parameter is invariant to threshold choice \citep{coles2001introduction}. Point estimates and $95\%$ confidence intervals for $\xi$ were obtained using exceedances of each candidate threshold, accounting for the rounding of observations. The confidence intervals for $\xi$ overlap for all candidate thresholds above 0.275$\text{M}_{\text{L}}$, and so by the parameter stability method any greater threshold is also valid. The thresholds chosen by our proposed method are therefore consistent with the parameter stability method, but are preferable in that the selected thresholds are not below the true level. Our proposed selection method is also more general; it allows comparison of many non-constant thresholds without the need for subjective and time-consuming interpretation of parameter stability plots. 

\begin{figure}
    \centering
    \includegraphics[width = 0.4 \textwidth, page = 1]{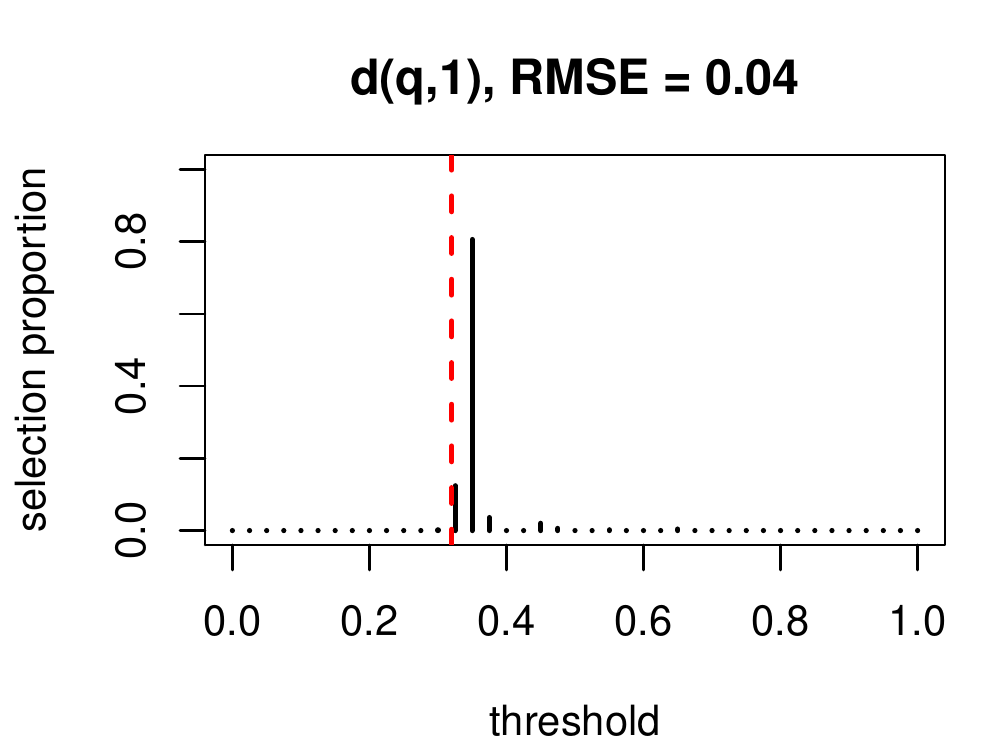}
    \qquad
    \includegraphics[width = 0.4\textwidth, page = 2]{figures/flat_threshold_sim/selected_threshold_summary_individual.pdf}
    \caption{Sampling distribution of threshold selection methods for quantile-based metrics over 500 simulated catalogues with constant threshold and hard censoring. The true threshold is shown by a dashed red line and the root mean squared error (RMSE) for each method is given in plot titles.}
    \label{fig:flat_threshold_selection_summary}
\end{figure}

Figure~\ref{fig:flat_threshold_selection_summary} presents the sampling distribution and RMSE of the thresholds selected from the candidate set by each of the QQ-based metrics over 500 replicated datasets, simulated as previously described. The thresholds chosen by the PP-based metrics are shown in Figure~E.2 of the Supplementary Materials \citep{varty2021inference} and are frequently much higher than the true value, resulting in higher RMSE values of 0.34 for $d(p,1)$ and 0.12 for $d(p,2)$. The metric $d(q,1)$ has the lowest RMSE and so appears to be the best of the proposed metrics in this case. All metrics have a tendency to overestimate the threshold value; this is likely to be attributable to the hard censoring process. We therefore also consider the performance of each metric using catalogues with phased censoring.

\subsection{Constant threshold, phased censoring} 
To assess the performance of each metric on simulated catalogues with phased censoring, we consider the thresholds selected by each metric for each of 500 simulated catalogues. For each catalogue, 2400 i.i.d. GPD exceedances of 0$\text{M}_{\text{L}}$ were simulated. Each exceedance was retained with probability $\alpha(y_i,v_i)$, as defined in Section~\ref{sec:simulation_study_overview} with $v(\tau) = 0.32 \text{M}_{\text{L}}$ and $\lambda = 7$. This combination of simulated catalogue size and censoring parameter gave an average catalogue size of 1500 recorded values, similar to those in Section~\ref{sec:constant_threshold_hard_censoring}.  

The resulting RMSEs in threshold selection over these 500 catalogues were: 0.06 for $d(q,1)$, 0.08 for $d(q,2)$, 0.35 for $d(p,1)$, and 0.12 for $d(p,2)$. For all metrics the RMSE is slightly increased compared to hard censoring, as threshold selection is made more difficult by the retention of some events that are truly below the threshold. As with hard censoring, the metrics $d(p,1)$ and $d(p,2)$ were prone to selecting conservative threshold values and $d(q,1)$ resulted in the lowest RMSE. Unlike for hard censoring, the sampling distributions of selected thresholds now cover the true threshold values, this is shown in Figure~E.4 of the Supplementary Materials \citep{varty2021inference}. Similar selection properties for each metric were seen when considering more complex threshold forms and so further exposition is limited to the metric $d(q,1)$, and we subsequently refer to $d = d(q,1)$.  

\subsection{Non-constant threshold selection}
\label{sec:sim_stutdy_non-constant_threshold}
\begin{figure}[hbt!]
    \centering
    \includegraphics[width = 0.4\textwidth]{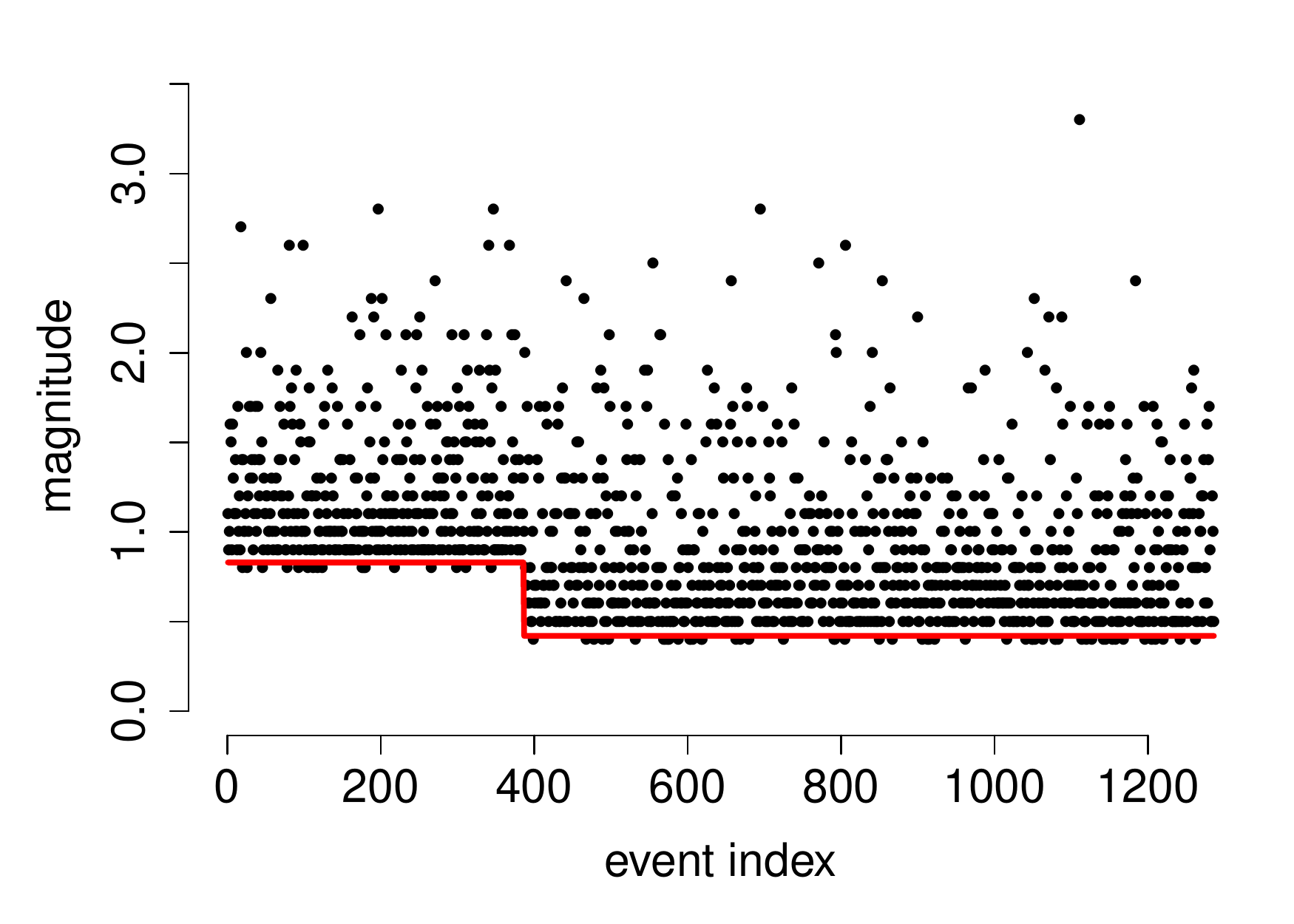}
    \qquad
    \includegraphics[width = 0.4\textwidth]{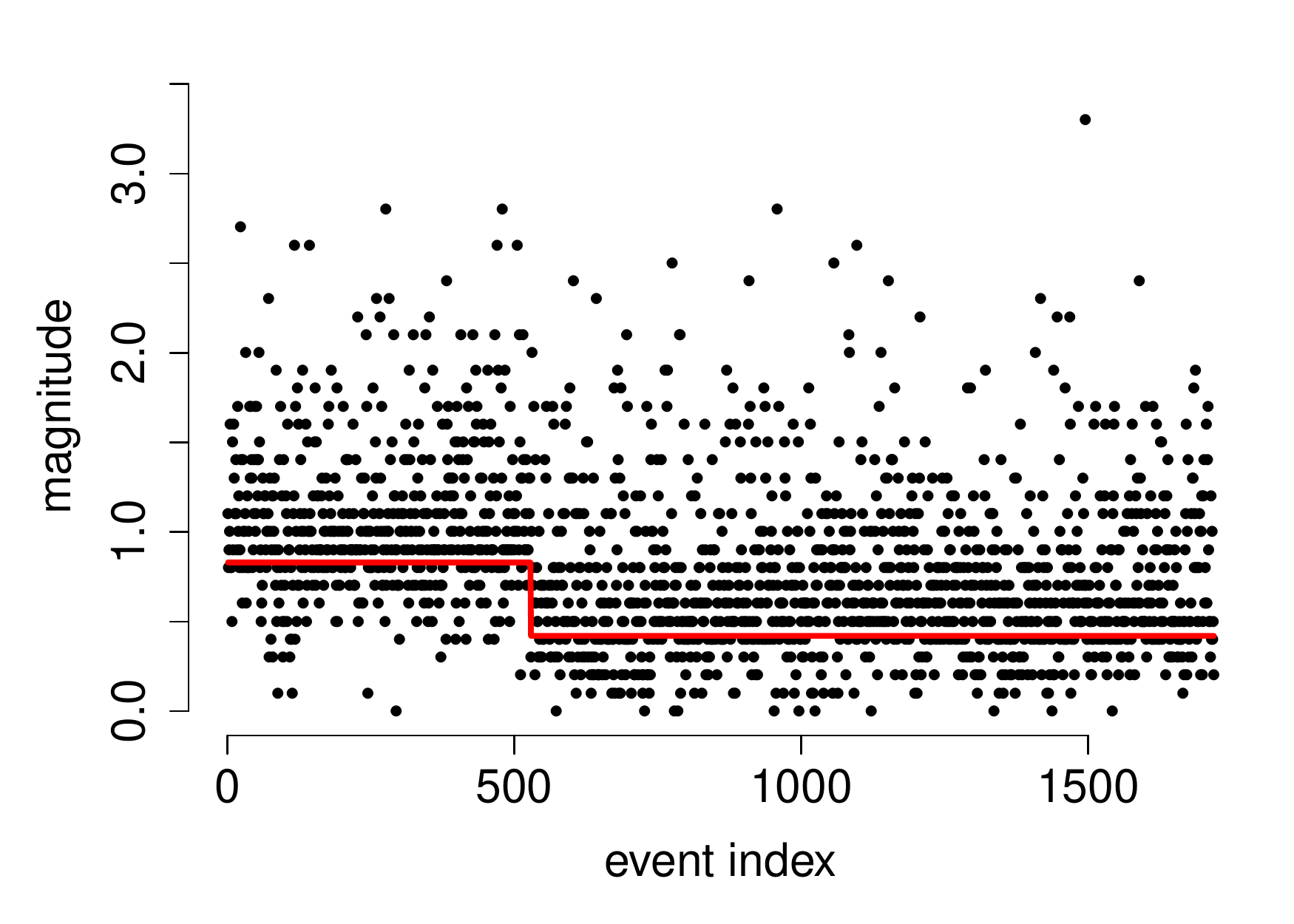}
    \caption{Example simulated catalogues with hard censoring [left] and phased censoring [right] for stepped thresholds of $(v^{(1)},v^{(2)})$ = (0.83,0.42), shown as a red line,  and phasing parameter $\lambda = 7$.}
    \label{fig:stepped_cat_examples}
\end{figure}

Here catalogues are simulated by generating 4000 i.i.d GPD exceedances of $0\text{M}_{\text{L}}$ and censoring (either hard or phased) based on a threshold with $v(\tau) =  0.83\text{M}_{\text{L}} $ for $0<\tau \leq 2000$ and $v(\tau) = 0.42\text{M}_{\text{L}}$ for $2000< \tau \leq 4000$, see Figure~\ref{fig:stepped_cat_examples} where $\lambda = 7$.

We considered threshold selection behaviour over 500 earthquake catalogues simulated using the above change-point threshold for each of hard and phased censoring. Note that the number of retained events and the threshold change location $\tau^*$ within these will vary between simulations because they both depend on the simulated event magnitudes and on how many of these are retained. However, in each case the true value of $\tau^*$ is known. 

\begin{figure}[hbt]
    \centering
    \includegraphics[width = 0.8\textwidth, page = 2]{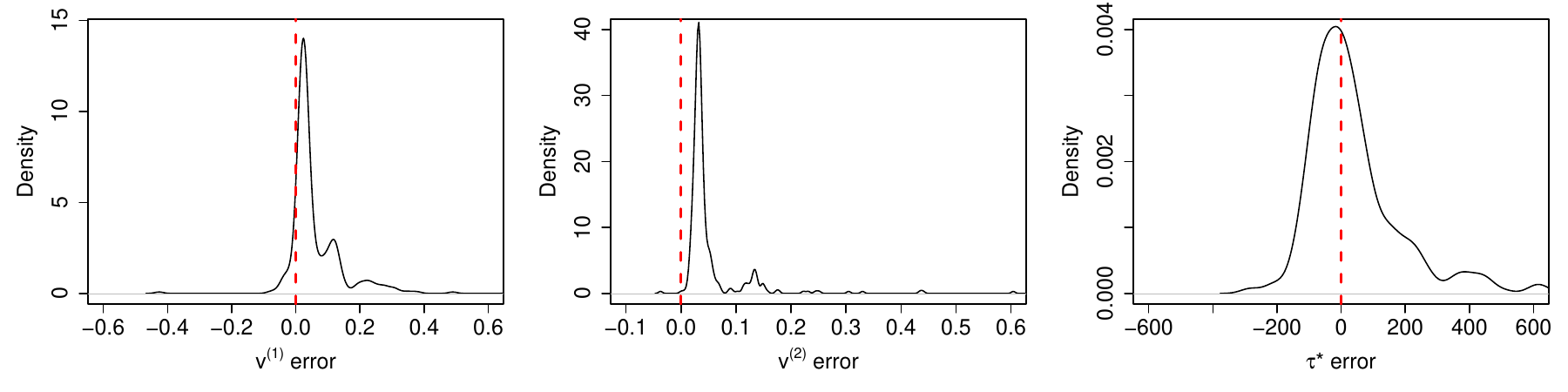}
     \includegraphics[width = 0.8\textwidth, page = 2]{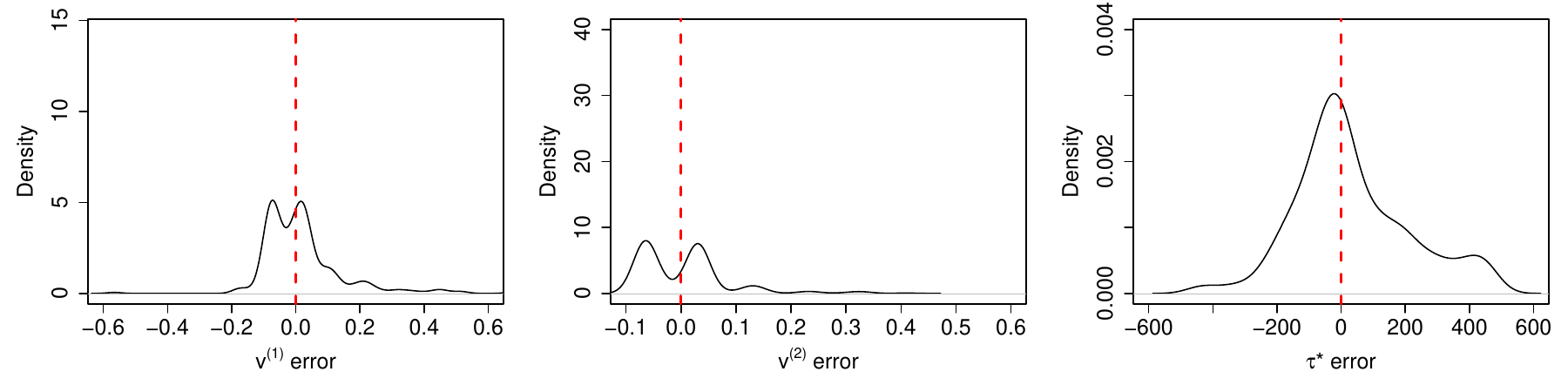} 
     
    \caption{Marginal sampling distributions of errors in the selected values of $v^{(1)}$ (left),  $v^{(2)}$ (center) and $\tau^*$ (right) for 500 simulated catalogues with change-point type thresholds and hard (top row) or phased (bottom row) censoring. }
    \label{fig:changepoint_errors}
\end{figure}

For each simulated catalogue we selected a threshold of the form $v(\tau) = v^{(1)}$ for $0 < \tau \leq \tau^*$ and $v(\tau) = v^{(2)}$ for $\tau^* < \tau < \tau_{\text{max}}$, where the threshold parameters $(v^{(1)}, v^{(2)}, \tau^*)$ are unknown. Threshold parameters were selected using the Bayesian optimisation method of Section~\ref{sec:minimisation_procedure} to minimise the metric $d$. The sampling distribution of the errors in the selected threshold parameters are shown in Figure~\ref{fig:changepoint_errors}, where it can be seen that our threshold selection method regularly recovers the non-constant modelling threshold to within $\delta/2$ of it true value. 

Specific findings vary by censoring type. For hard censoring, as would be expected, the threshold levels $v^{(1)}$ and $v^{(2)}$ are rarely selected to be below the true values. The error distribution of $\tau^*$ has, in both cases, a mode close to 0 but with large variance. As expected, the sampling variability of the error in each parameter is larger for phased censoring than for hard censoring, though it is reassuring to see that the distributions of selected threshold parameters are now centered on the true values. This demonstrates that the tendency to select threshold values too high for catalogues with hard censoring is a consequence of the censoring mechanism, not a bias in the selection method. 
 
\section{Application to Groningen earthquakes} 
\label{sec:groningen_application}
\subsection{Validating data model for Groningen catalogue} \label{sec:groningen_application_validating_rounded_gpd}
We compare GPD and exponential models for Groningen earthquake magnitudes. \citet{rohrbeck2018extreme} and \citet{marzocchi2019how} demonstrated the importance of acknowledging rounding of observations, and so this is accounted for within the inference for both models. We focus on earthquakes exceeding the constant conservative threshold of 1.45$\text{M}_\text{L}$, subsequently referred to as $v_C$. This is the magnitude of completion stated by the KNMI \citep{dost2012monitoring}, adjusted to account for rounding.
 
Both the GPD and exponential models assume that magnitudes are i.i.d.; this is supported by our exploratory analysis of the Groningen catalogue in Section~A of the supplementary materials \citep{varty2021inference}. The two models may be compared by considering the sampling distribution of the estimated shape parameter under a GPD model, because the exponential model is a special case of the GPD where $\xi = 0$. Fitting a GPD to the 311 exceedances of $v_C$ leads to point estimates of $(\hat\sigma_{1.45}, \hat\xi) = (0.448, -0.018)$ with respective $95\%$ bootstrap confidence intervals of $(0.399,0.501)$ and $(-0.147,0.086)$. Since the confidence interval for $\xi$ covers 0, the exponential model cannot be discounted at the $5\%$ significance level using only exceedances of $v_C$. A second method of comparison is to fit both an exponential and GPD model to exceedances of $v_C$ and, appealing to the asymptotic distribution of the MLE, perform a likelihood ratio test. This produces a likelihood ratio of 1.04 and associated $p$-value of 0.214, leading us to draw the same conclusion in both comparisons: that there is insufficient evidence to conclude that the Groningen magnitudes deviate from the Gutenberg-Richter law when using only exceedances of $v_C$.

However, if an exponential magnitude model is assumed then the uncertainty about $\xi$ is ignored. This has the effect of dramatically, but artificially, narrowing the confidence intervals on the estimated magnitude quantiles, as shown in Figure~E.3 of the Supplementary Materials \citep{varty2021inference}. The potential repercussions of ignoring this uncertainty are described in detail in \citet{coles2003anticipating}. A GPD model should therefore be used for the underlying magnitudes, to properly represent this uncertainty when selecting a modelling threshold for the Groningen gas field. 

If the rounding of observations had been ignored in the fitting of the GPD model, the point estimates of the GPD parameters would be $(\hat\sigma_{1.45}, \hat\xi) =  (0.453, -0.027)$ with respective standard errors of $(0.039, 0.066)$. The parameter estimates are not significantly different to those using the correct likelihood because the small number of threshold exceedances means that parameter uncertainty obscures the bias induced by neglecting to account for rounding.

\begin{figure}
    \centering
    \includegraphics[width = 0.3\textwidth, page = 1]{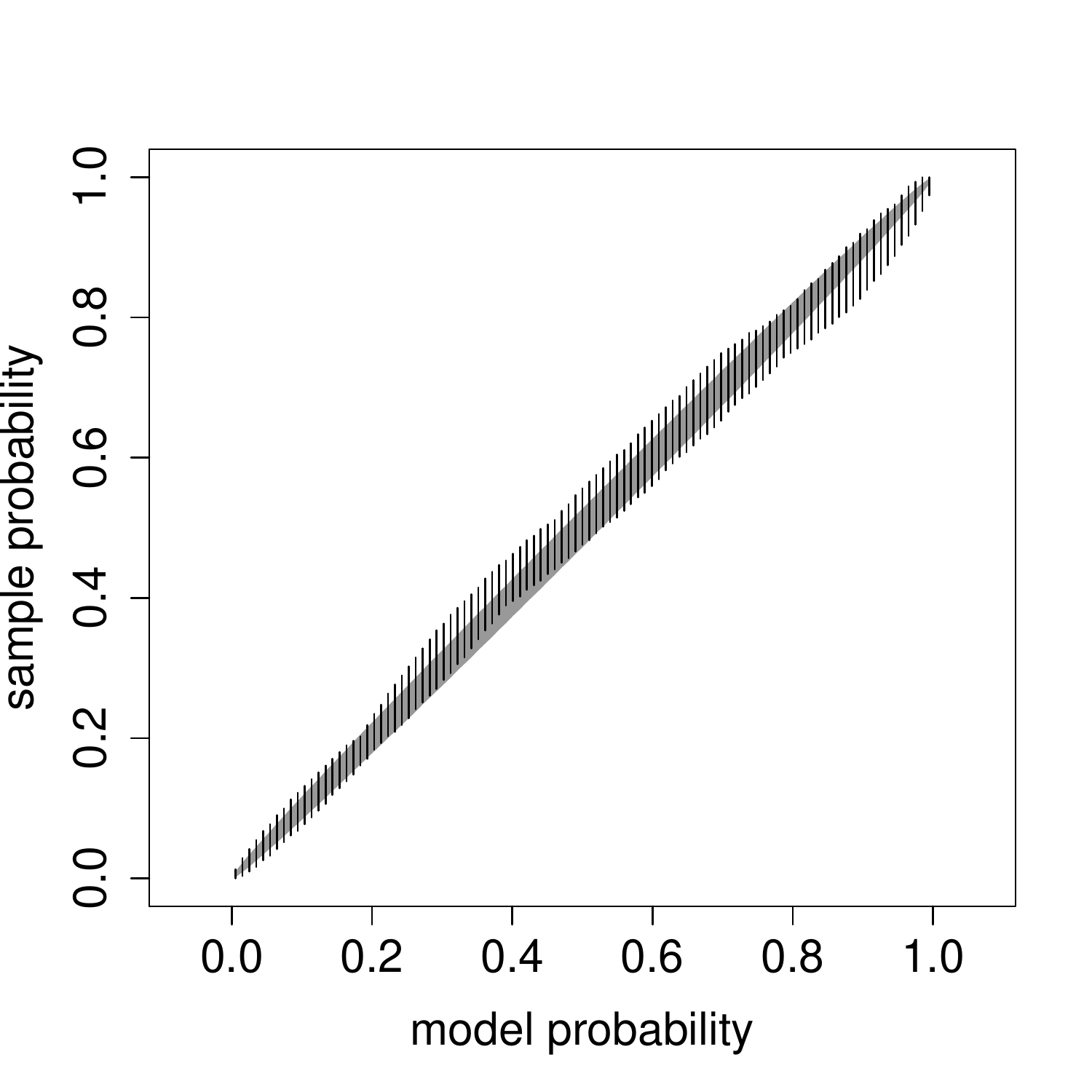}
    \qquad
    \includegraphics[width = 0.3\textwidth, page = 2]{figures/groningen_application/motivating_GPD_over_exponential/groningen_conservative_mod_PP_QQ.pdf}
    \caption{Modified PP (left) and QQ (right) plots for Groningen magnitudes exceeding 1.45$\text{M}_{\text{L}}$ under the GPD model. Grey regions show $95\%$ tolerance intervals while vertical lines show $95\%$ confidence intervals on sample probabilities / quantiles. All confidence intervals overlap with the associated tolerance intervals.  }
    \label{fig:groningen_mod_pp_qq_above_145}
\end{figure}

Finally, in Figure~\ref{fig:groningen_mod_pp_qq_above_145} we check that the fitted GPD model is consistent with the empirical distribution of exceedances of $1.45\text{M}_{\text{L}}$ through the use of the modified QQ and PP plots introduced in Section~\ref{sec:threshold_selection_graphical}. Since the tolerance intervals and confidence intervals overlap for both the sample quantiles and sample probabilities, we conclude that a GPD model is appropriate for Groningen earthquake rounded magnitudes exceeding $1.45 \text{M}_{\text{L}}$.

\subsection{Parametric threshold forms} 
\label{sec:groningen_application_threshold_selection} 
 Now we select thresholds of two parametric forms for the Groningen catalogue and explore the results of the subsequent inference. The first is a constant threshold, $v(\tau) = v$, where the level $v$ is to be chosen. This will allow us to assess the level of conservatism in the conventional modelling threshold where $v = 1.45 \text{M}_{\text{L}}$. The second form is a sigmoid-type threshold
    $ v(\tau) = v_{R} + (v_{L} - v_{R}) \Phi\left([{\mu - \tau}]/{\varsigma}\right),$
 with parameters $(v_{L}, v_{R}, \mu, \varsigma) \in \mathbb{R}^3 \times \mathbb{R}^+$ and where $\Phi$ is the standard Gaussian distribution function.
 This extends the idea of the change-point threshold to accommodate smooth change in the threshold value centred on $\mu$. The threshold parameters may be interpreted as follows. The left and right asymptotic levels of the threshold are given by $v_L$ and $v_R$, $\mu$ is the index-time at which the threshold takes the value  $(v_{L} + v_{R})/2$, and $\varsigma$ controls how rapidly the threshold changes about $\mu$, with $\varsigma \rightarrow 0$ corresponding to a step change. In the context of the Groningen catalogue we expect that $v_R < v_L$. 
\subsection{Threshold selection} 
\subsubsection{Constant threshold}
\begin{figure}
    \centering
    \includegraphics[width = 0.4\textwidth]{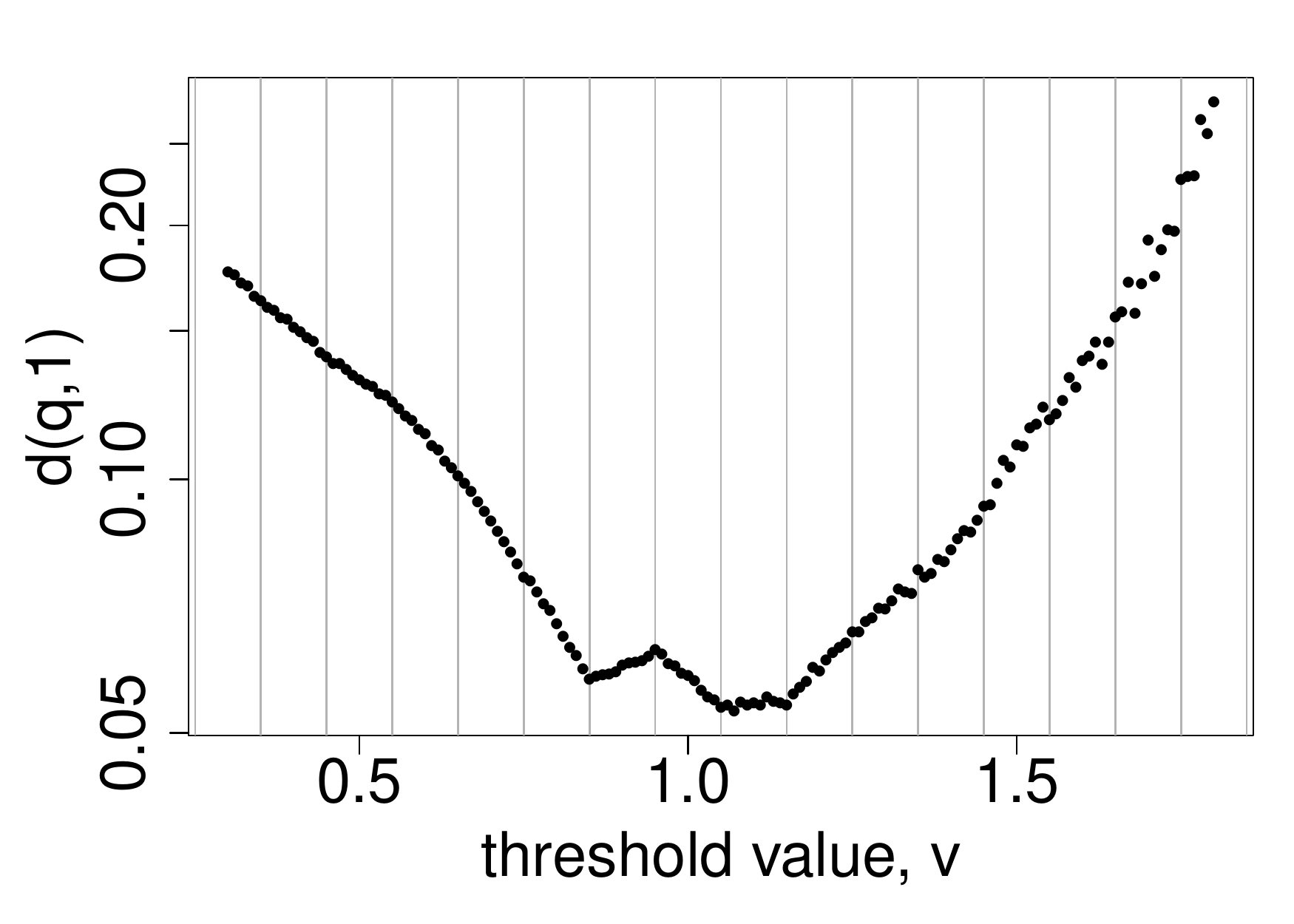}
    \qquad
    \includegraphics[width = 0.4\textwidth]{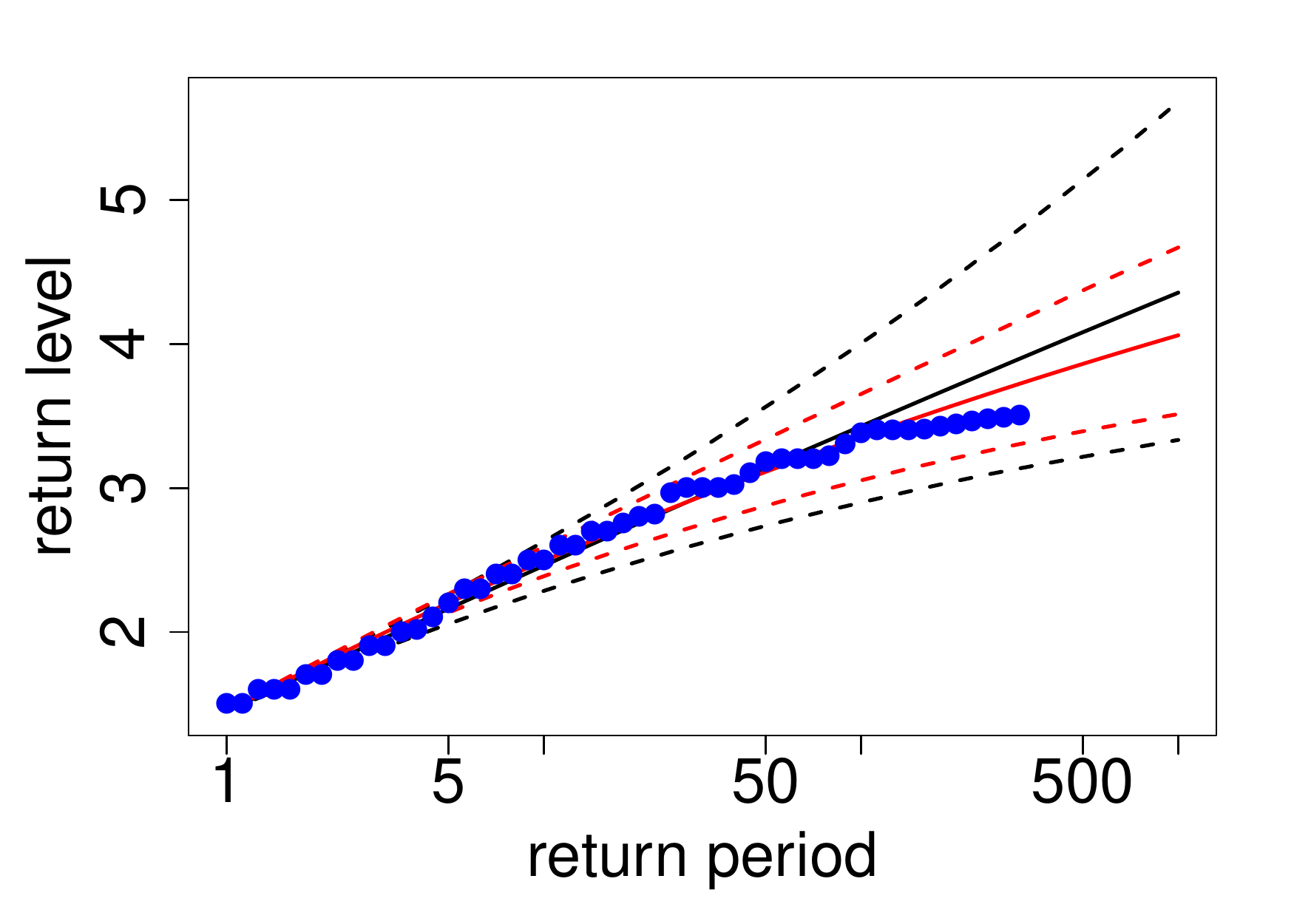}
    \caption{[Left] Grid search to minimise $d(q,1)$ over threshold values $v$. Metric values are shown on log-scale and vertical lines mark the edges of magnitude rounding intervals. [Right] Point estimates (solid lines) and $95\%$ confidence intervals (dashed lines) for the conditional return levels for exceedances of $1.45 \text{M}_{\text{L}}$, using the conservative (black) and selected thresholds (red). Sample conditional return levels are shown in blue.}
    \label{fig:groningen_flat_selection_metrics_values}
\end{figure}

A grid search was used to find the flat threshold that minimises the metric $d$, as shown in Figure~\ref{fig:groningen_flat_selection_metrics_values}. There are two local minima at $v = 0.85\text{M}_{\text{L}}$ and $v = 1.07\text{M}_{\text{L}}$, the latter being the global minimum. For thresholds greater than $1.07\text{M}_{\text{L}}$, including the conservative threshold of $1.45\text{M}_{\text{L}}$, the metric values are increasing as not all viable data are utilised. For thresholds less than $0.85\text{M}_{\text{L}}$ the metric also increases as the validity of the tail model breaks down. The small peak between these minima is likely attributable to the reduction of the $m_c$ over time. In Figure~\ref{fig:groningen_catalogue} we saw that fewer small magnitude events are censored at later times. The minimum at $1.07\text{M}_{\text{L}}$ uses less data to achieve good distributional agreement for the entire period, while the minimum at $0.85\text{M}_{\text{L}}$ compromises on the distributional agreement at early times to retain a larger proportion of the data. As the threshold is lowered between magnitudes $0.95\text{M}_{\text{L}}$ and $0.85\text{M}_{\text{L}}$, enough additional data are added to more than compensate for the reduced goodness-of-fit in the early part of the observation period and so the metric value reduces. Since the global minimum corresponds to the more conservative threshold, we select $1.07\text{M}_{\text{L}}$ as our constant modelling threshold.

\subsubsection{Sigmoid threshold}
Bayesian optimisation was used to find the sigmoid threshold parameters $(v_{L}, v_{R}, \mu, \varsigma)$ that minimise the metric $d$, where the search space was constrained to the region $[0.4,1.7]^2 \times [200,1100] \times [1,500]$. For an initial set of 20 randomly selected threshold parameter combinations, $d$ was evaluated. A further fixed budget of 100 metric evaluations was allocated and the thresholds with the smallest metric value retained for further inspection. To assess the sensitivity of the selected threshold to the set of initial evaluation points, this was repeated for five initial parameter combination sets. 

\begin{figure}
    \centering
    \includegraphics[width = 0.32\textwidth]{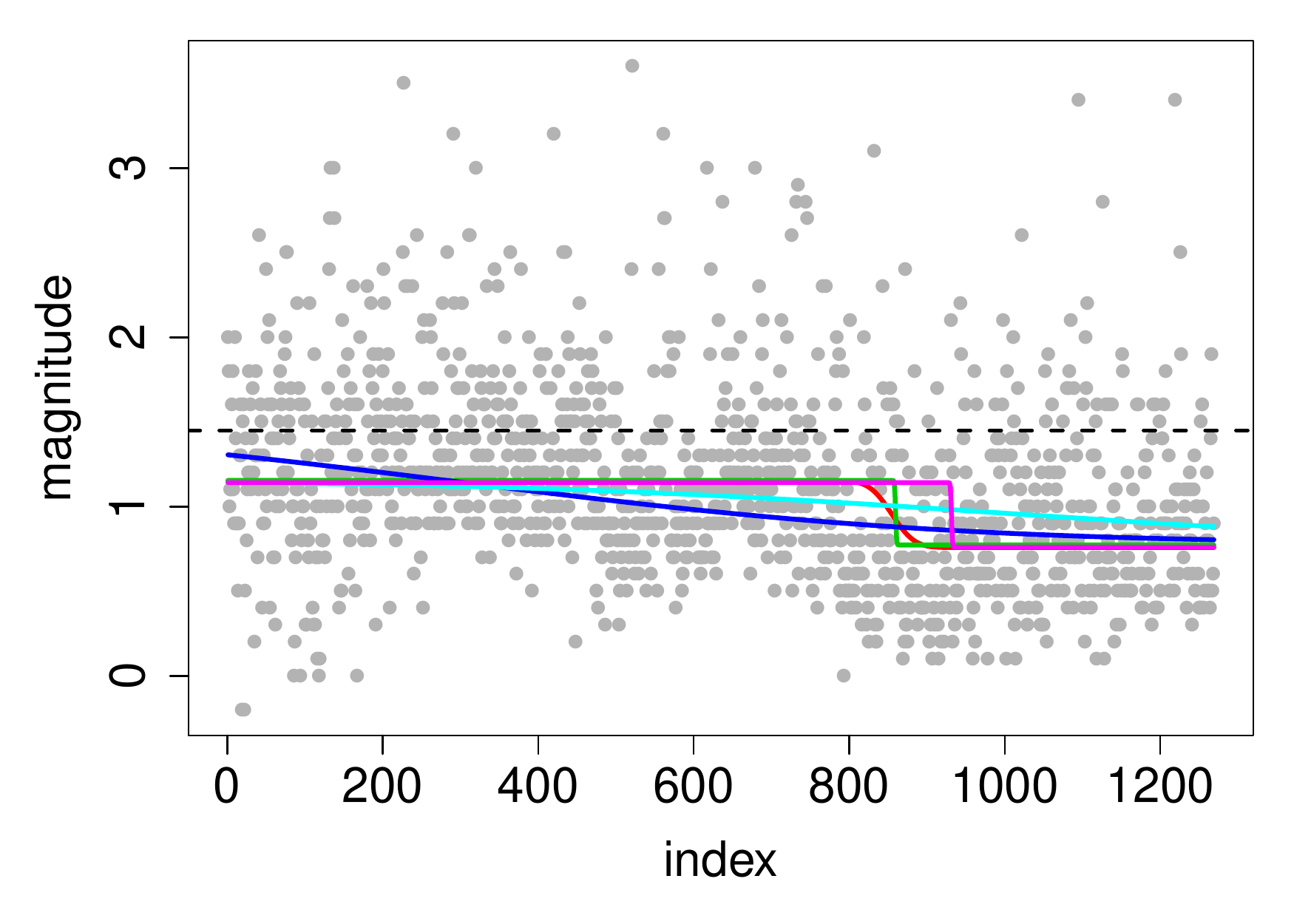} 
    \includegraphics[width = 0.32\textwidth]{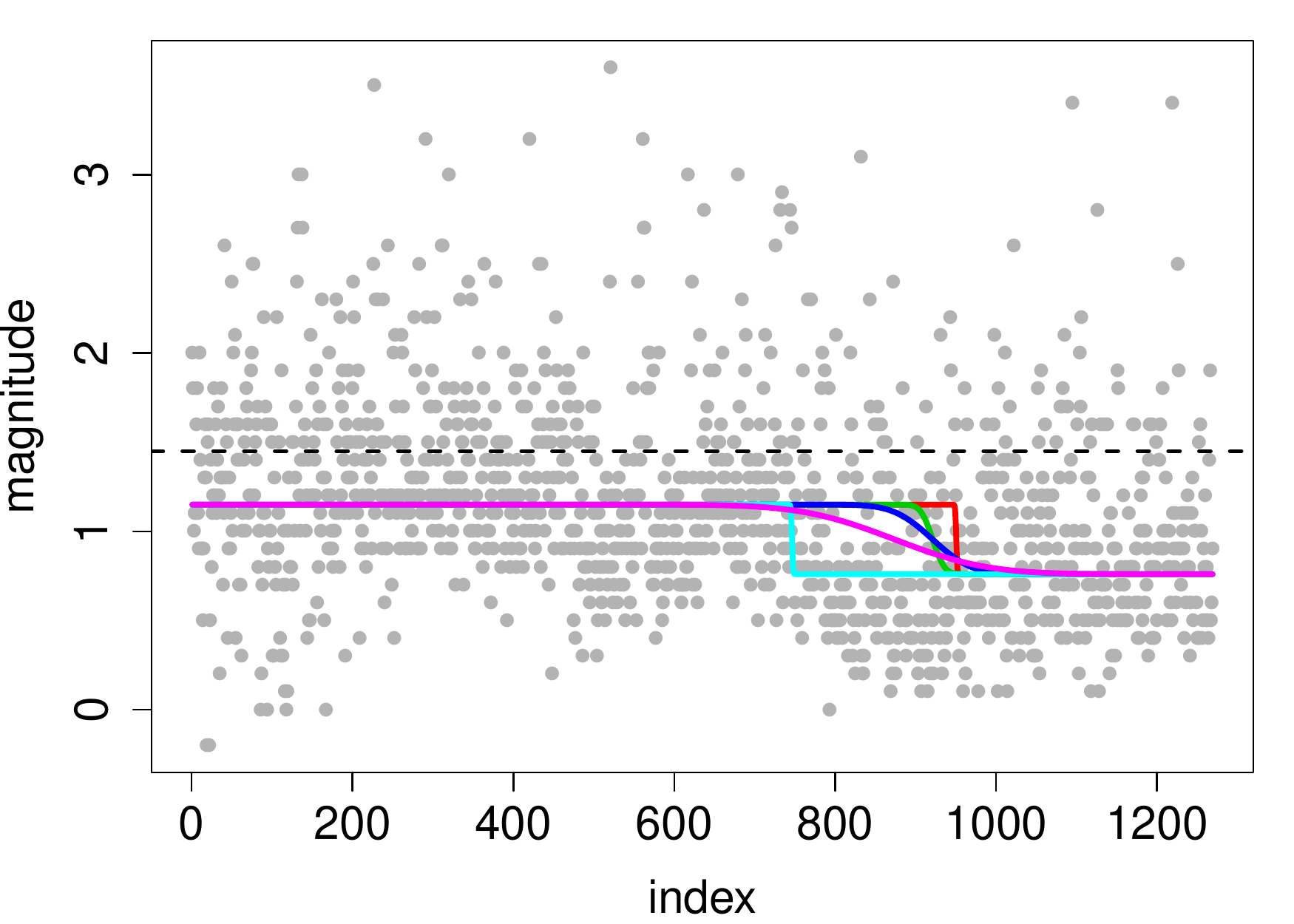}
    \includegraphics[width = 0.32\textwidth, page = 2]{figures/groningen_application/sigmoid_threshold_selection/BO_selected_thresholds_restricted.pdf}
    \caption{ Selected sigmoid thresholds using Bayesian optimisation from 5 random initial parameter sets. [left] Optimising over all thresholds parameters. [centre, right] Optimising over $(\mu, \varsigma)$ and fixing $(v_{L}, v_{R})$ = $(1.15,0.76)$ on index- (centre) and natural- (right) timescales. Colours are comparable only between centre and right plots. Dashed horizontal lines show the conservative threshold value. Important dates relating to the development of the Groningen seismic detection network are shown as vertical lines: (A) development begins, (B) first additional sensors activated, (C) upgrade complete.}
    \label{fig:groningen_selected_sigmoid_thresholds}
\end{figure}

The thresholds with the lowest values of $d$ based on each initialisation are shown in Figure~\ref{fig:groningen_selected_sigmoid_thresholds} (left). The selected threshold values at the ends of the observation interval appear to be stable across initialisation, but the transition between these levels is not.  Further investigation supports the stability of the end levels; the blue and turquoise thresholds have significantly greater metric values than the other thresholds, suggesting that these initialisations had too few evaluations to explore beyond a local minimum. These conclusions are consistent with the simulation study of Section~\ref{sec:sim_stutdy_non-constant_threshold}, illustrating that threshold levels are more easily estimated than the change between those levels. 

A second Bayesian optimisation was performed, fixing the end levels of the sigmoid threshold to the those shared by the best performing thresholds in the previous optimisation,  namely $(v_{L}, v_{R})$ = $(1.15,0.76)$. This reduces the dimension of the parameter space and simplifies the optimisation task. Using the same procedure as for the unconstrained optimisation, the resulting selected thresholds from each initialisation are shown in Figure~\ref{fig:groningen_selected_sigmoid_thresholds} (centre). Upon repeated Monte Carlo evaluation of the metric value for each of these thresholds, there is insufficient evidence to select one over the others. When transformed onto the natural time scale, as shown in Figure~\ref{fig:groningen_selected_sigmoid_thresholds} (right), the selected thresholds are all consistent with the known dates at which sensor installation occurred. This shows that from the earthquake catalogue alone our method is able to detect the starting and ending threshold levels and the period in which it changed. However, we cannot identify precisely the way in which the threshold changed during the installation period. This is not a major setback, since between the most and least conservative of the chosen thresholds (turquoise and red in the centre and right panels of Figure~\ref{fig:groningen_selected_sigmoid_thresholds}) the expected number of observations above the threshold differs by only 50 earthquakes. We fitted the GPD model using each of these five threshold functions, reaching similar conclusions, and so present further results for only the turquoise threshold. 

\subsubsection{Threshold comparison} 
We compare the conservative, selected constant, and selected sigmoid thresholds, which are referred to as $\hat v_C, \hat v$ and $\hat v_S$ respectively. Comparisons are made on: the expected metric value, the number of events used to fit the GPD model, the estimated GPD parameter values, and the estimated return levels.

Metric evaluations are subject to Monte Carlo noise and so the metric value was evaluated 100 times for each threshold. The mean metric value and $95\%$ Monte Carlo noise intervals were calculated to be: 0.091 (0.088, 0.096) for $\hat v_C$, 0.054 (0.053, 0.055) for $\hat v$, and 0.041 (0.039, 0.043) for $\hat v_S$. This suggests that the model fit using $\hat v_S$ fits the observed data best, with $\hat v$ being preferred over $\hat v_C$. 
These improvements in model fit may be attributable to the increased data usage of the selected thresholds. The threshold $\hat v_C$ is at the edge of a rounding interval and so utilises 311 threshold exceedances in the resulting model. For thresholds $\hat v$ and $\hat v_S$, the rounding of magnitudes means that the exact number of exceedances is unknown. The expected number of exceedances under the fitted magnitude models are 629 and 702 for $\hat v$ and $\hat v_S$, respectively. By using either of the selected thresholds, we have more than doubled the size of usable catalogue as compared to the conservative threshold. 

\begin{figure}
    \centering
    \includegraphics[width = 0.4\textwidth]{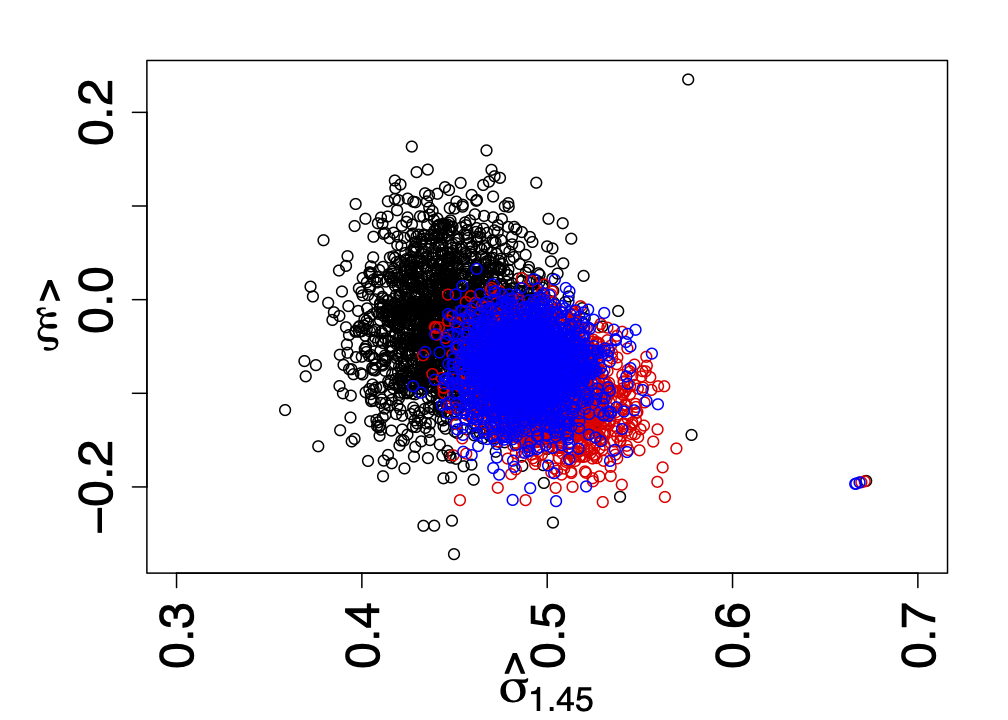}
    \qquad
    \includegraphics[width = 0.4\textwidth]{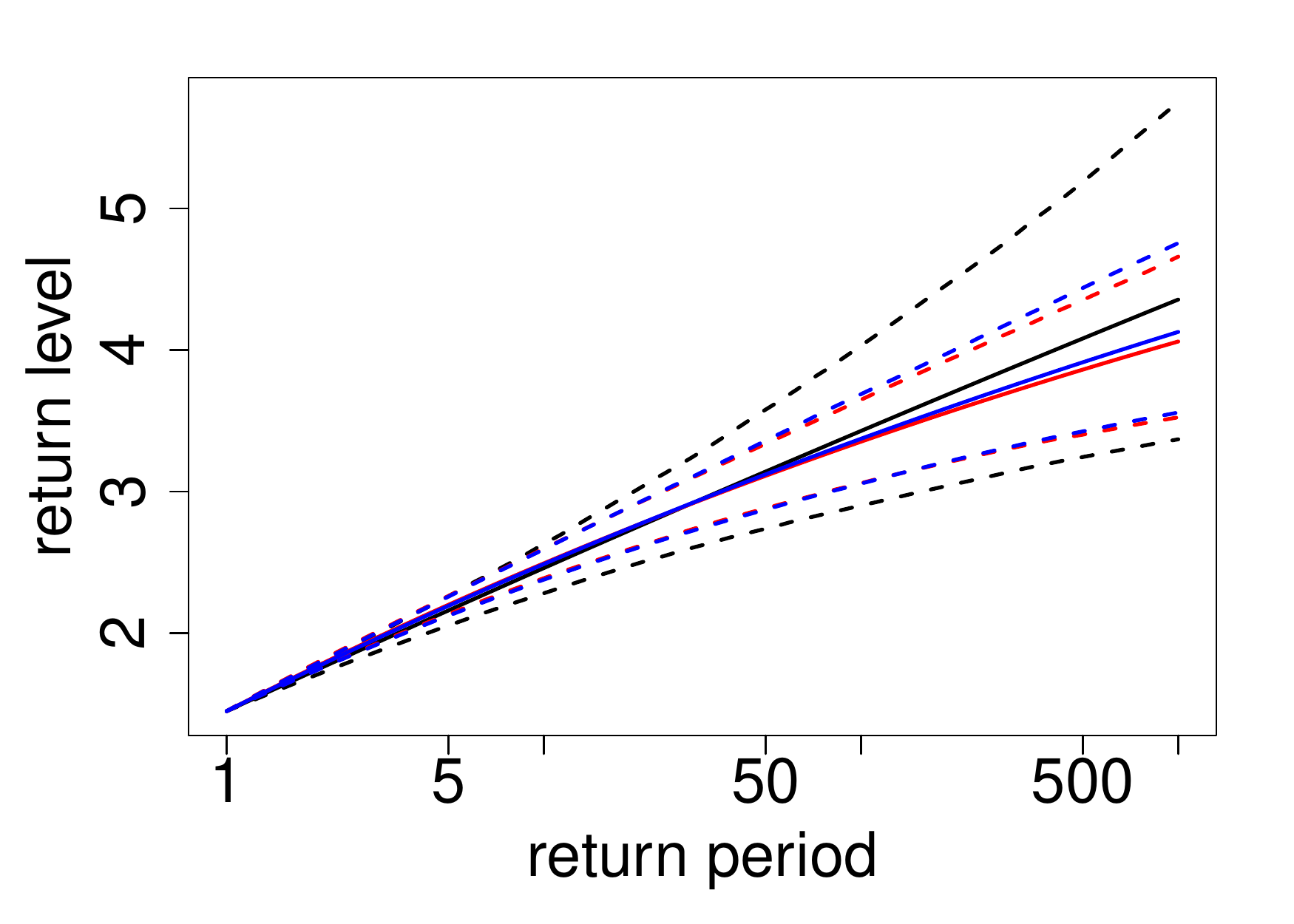}
    \caption{Bootstrap GPD parameter estimates based on exceedances of the conservative (black), flat (red) and sigmoid (blue) thresholds [left]. Estimated return levels in $\text{M}_{\text{L}}$ and $95\%$ confidence intervals for magnitudes exceeding $1.45\text{M}_{\text{L}}$ [right]. }
    \label{fig:groningen_threshold_comparison}
\end{figure}

Figure~\ref{fig:groningen_threshold_comparison} (left) shows the estimated parameter values under the fitted GPD model using each threshold. The uncertainty in both parameters is reduced when using $\hat v$ rather than $\hat v_C$, and further reduced when using $\hat v_S$. To give a sense of scale in this reduction, we can calculate the number of additional exceedances of $\hat v_C$ to which they are equivalent, under the assumption that the standard error of parameter estimates scales with exceedance count $n$ as $n^{-1/2}$.  In doing this, the additional $318$ and $391$ small magnitude earthquakes included by, respectively, using $\hat v$ or $\hat v_S$ are equivalent to $363$ or $509$ additional events above $\hat v_C$. Therefore, point-for-point, the small magnitude earthquakes are at least as valuable as additional data above $v_C$ for parameter estimation. 

When modelling exceedances of $\hat v$ or $\hat v_S$ the respective point estimates and $95\%$ confidence intervals for the shape parameter are -0.084 (-0.168, -0.017) and -0.069 (-0.144, -0.008). Using exceedances of $\hat v$ or $\hat v_S$ leads to only $0.5\%$ or $1.5\%$ of the sampling distribution for $\hat\xi$ being above 0. This provides empirical evidence that the Groningen magnitude distribution has a finite upper endpoint, unlike the conventional Gutenberg Richter magnitude model. This dramatic conclusion could not be reached using the smaller dataset exceeding $\hat v_C$, where $33\%$ of the sampling distribution for $\hat \xi$ lay above 0. 

Similar conclusions can be reached by using likelihood ratio tests to compare Exponential and GPD models for exceedances each of $v_C$, $\hat v$ and $\hat v_S$; the respective $p$-values are 0.78, 0.046, and 0.064. By using more of the available data, we have increased our ability to discern between an exponential model and the observed magnitude distribution. The conclusions that can be drawn from this test are in agreement with, but are less strong than, those of the previous comparison: a Gutenberg Richter magnitude model is likely inferior to a GPD. The discrepancy in conclusion strength between the two comparisons is likely due to the asymptotic assumptions of the likelihood ratio test not being met by our finite sample size.    

The estimated conditional return levels above $1.45 \text{M}_{\text{L}}$ are shown using each threshold in Figure~\ref{fig:groningen_threshold_comparison} (right). The estimated return levels are similar when using $\hat v$ and $\hat v_S$, but confidence intervals for large return periods are narrower when using $\hat v_S$. In either case, the return levels have both smaller point estimates and uncertainties by using our threshold selection method than when using the conservative threshold. This is an important finding when deciding what measures to take when designing or retrofitting earthquake defences for buildings.

\section{Discussion / Conclusion} \label{sec:conclusion}
This paper introduced a principled method to select a time-varying modelling threshold for an extreme value analysis. The effectiveness and value of using this method to include additional, less extreme events in the analysis were demonstrated through simulation studies. Although the method was developed in the context of earthquake catalogues, and to accommodate the additional challenges to inference that these pose, the core of our method is applicable to extreme value threshold selection more generally and we anticipate it having a much broader impact.

Using the new threshold selection method, we have been able to identify the period in which the Groningen sensor network was being improved by using the earthquake catalogue alone. Our threshold selection method more than doubled the usable size of the Groningen earthquake catalogue compared to using the conservative threshold given by the KNMI, whilst also improving model fit. This has several important implications beyond the direct improvement to statistical inference.

The use of these additional small magnitude earthquakes leads to greater precision in the estimates of high magnitude quantiles, which is potentially a huge benefit by reducing the cost of designing, constructing or retrofitting earthquake defences. Following threshold selection, a Bayesian modelling approach would allow quantile uncertainty to be included naturally when designing defences against natural hazards \citep{coles1996bayesian, fawcett2018bayesian, jonathan2020uncertainties} and estimates with greater precision can reduce the cost required to provide protection with equivalent confidence. The gain we have made in the efficiency of statistical inference can be translated to a tangible economic benefit of using the additional data recorded by improving the censor network.
The more efficient use of the available data has allowed us to conclude, for the first time based on empirical evidence alone, that Groningen earthquake magnitudes are likely to have a light-tailed distribution. Using the conservative threshold level this conclusion could only have been achieved by waiting many years to observe additional large magnitude earthquakes. 
Finally, using a less conservative modelling threshold provides a return on the substantial investment into the earthquake detection network around the Groningen gas field. When a non-constant threshold is selected, the added value of the network improvements is exploited fully and the subsequent modelling threshold can also offer insights into the reduction of $m_c$ over time.

A limitation of the work is that the computational effort required to select a modelling threshold is relatively high. We do not view this as a large drawback since threshold selection must be performed only once through the modelling process. An area for further development would be to investigate alternative, exact methods to optimise the expected selection metric over the threshold parameters. One possible extension to our approach would be to adapt the data model to account for magnitude measurement error causing events to be recorded within incorrect rounding intervals. Another, more ambitious, extension might consider a selection of spatio-temporal threshold function to describe spatial variability as well as the temporal evolution of event detection. Finally, an extensive comparison of our proposed and standard extreme value threshold selection methods would be a valuable piece of further work, given its critical importance in extreme value methods.

\begin{appendix}
%
\end{appendix}

 \section*{Acknowledgements}
%
\textit{
This paper is based on work completed while Zak Varty was part of the EPSRC funded STOR-i centre for doctoral training (EP/L015692/1), with part-funding from Shell Research Ltd.}
 

\begin{supplement} 
\stitle{Supplement:} 
\sdescription{Further detail is given on the methods introduced in the main text. Additional plots are provided in support of our simulation studies and the analysis of Groningen earthquakes.}
\end{supplement}


\bibliographystyle{imsart-nameyear} 
\bibliography{references}       


\end{document}


\begin{frontmatter}
\title{Supplementary materials for "Inference for extreme earthquake magnitudes accounting for a time-varying measurement process"}
\runtitle{Supplementary materials}

\begin{aug}
\author[A]{\fnms{Zak} \snm{Varty}\ead[label=e1]{z.varty@lancaster.ac.uk}},
\author[A]{\fnms{Jonathan A.} \snm{Tawn}\ead[label=e2,mark]{j.tawn@lancaster.ac.uk}}
\author[A]{\fnms{Peter M.} \snm{Atkinson}\ead[label=e3,mark]{pma@lancaster.ac.uk}}
\and
\author[B]{\fnms{Stijn} \snm{Bierman}\ead[label=e4,mark]{stijn.bierman@shell.com}}
\address[A]{Lancaster University,
\printead{e1,e2,e3}}

\address[B]{Shell Global Solutions Netherlands,
\printead{e4}}
\end{aug}



\end{frontmatter}



\begin{appendix}
\section{Assessing i.i.d. assumption for Groningen earthquakes}
\subsection{Connection to main text} This appendix supports the claim made in Sections 2 and 6 of the main text that Groningen earthquakes exceeding $1.45 \text{M}_{\text{L}}$ may be modelled as independent and identically distributed. 

\subsection{Exploratory analysis}
Here we examine the validity of the assumption, common to both GPD and exponential models, that magnitudes are i.i.d. above 1.45$\text{M}_{\text{L}}$. In the case of continuous-valued data that are completely observed, this assumption implies that inter-arrival times of threshold exceedances should approximately follow an exponential distribution. Due to incomplete observation below $1.45 \text{M}_{\text{L}}$ and the transformed time scale, we instead consider the inter-arrival times of events exceeding magnitude $c \geq v_C$ measured in terms of the number of events with magnitudes between 1.45$\text{M}_{\text{L}}$ and $c$. If events are i.i.d. then these inter-arrival times are geometrically distributed. It is important to investigate a range of values for $c$, since lower values lead to more (shorter) observed interval lengths but these are more concentrated about 0 making assessment of the geometric distribution more difficult. This trade-off can seen by considering the edge-cases: if $c = 1.45 \text{M}_{\text{L}}$ then each interval is of length 0, and if $c$ is between the second and third largest observed magnitudes then there is a single observed interval.
 
\begin{figure}
     \centering
     \includegraphics[width = 0.3\textwidth, page = 2]{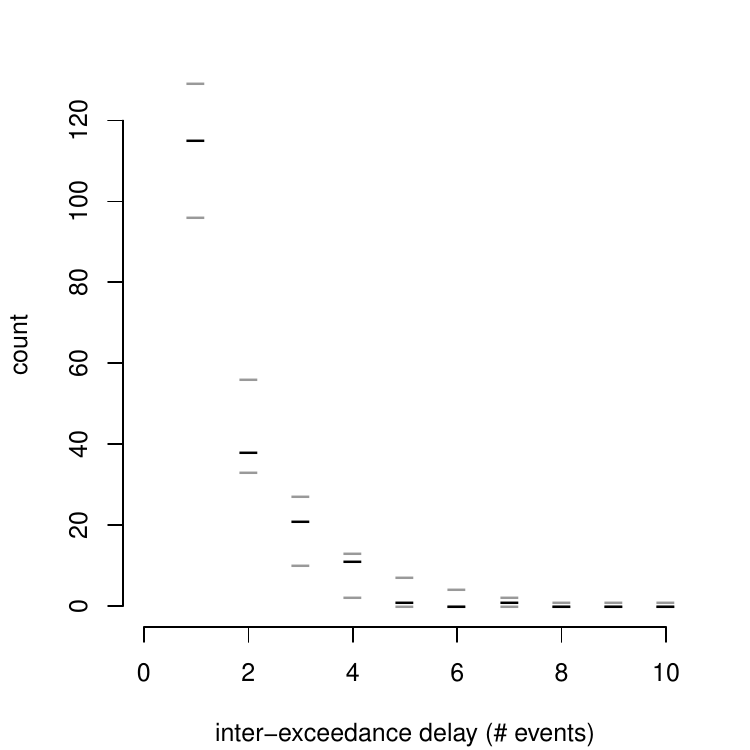}
     \includegraphics[width = 0.3\textwidth, page = 3]{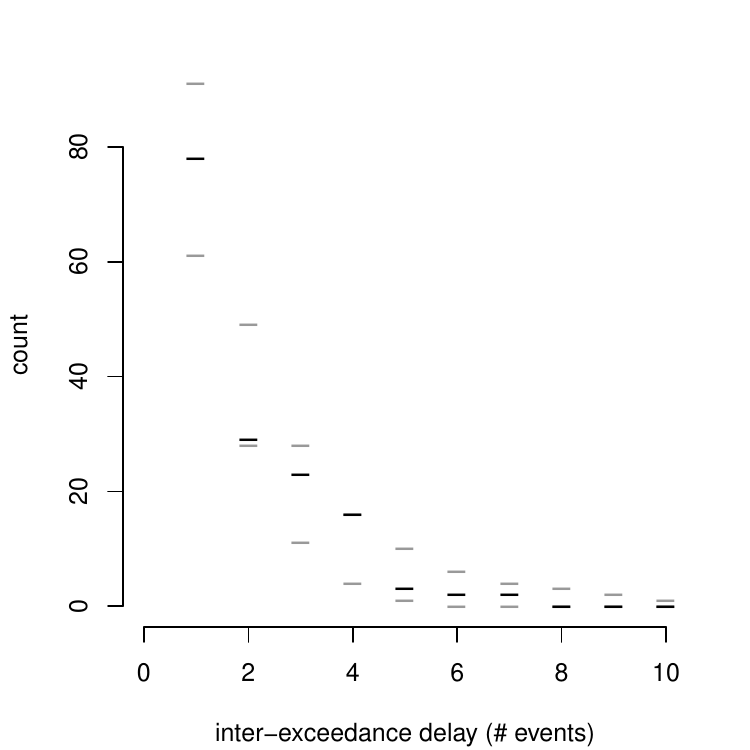}
     \includegraphics[width = 0.3\textwidth, page = 4]{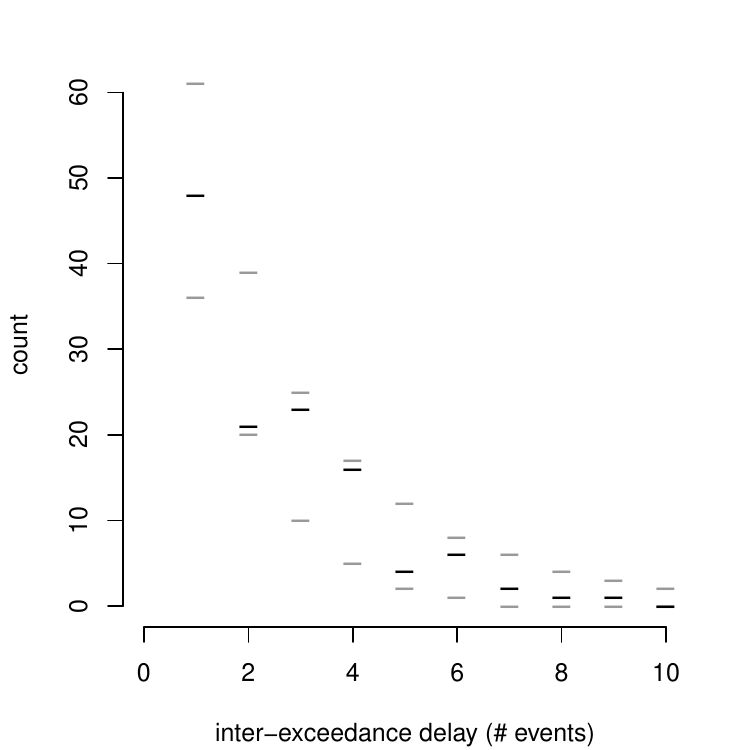}
     \caption{Frequency plots of the number of earthquakes exceeding 1.45$\text{M}_\text{L}$ that separate earthquakes exceeding exceeding 1.65$\text{M}_\text{L}$ (left), 1.75$\text{M}_\text{L}$ (centre), and 1.85$\text{M}_\text{L}$ (right) for the Groningnen earthquake catalogue. Observed frequencies (black lines) fall within the $95\%$ confidence intervals under the fitted models (grey lines).}
     \label{fig:groningen_rgpd_above_15}
 \end{figure}

 The empirical distributions of interval lengths in the Groningen catalogue are shown in the panels of Figure~\ref{fig:groningen_rgpd_above_15} for exceedances of $c = 1.65 \text{M}_\text{L}$, $1.75 \text{M}_\text{L}$, and $1.85 \text{M}_\text{L}$. These are consistent with the $95\%$ confidence interval for the fitted geometric distribution at each value of $c$. The same conclusion was found for $1.55\text{M}_\text{L} < c <  2.25\text{M}_\text{L}$. Events larger than this show mild evidence of clustering, but overall this suggests that it is reasonable to model magnitudes as i.i.d. above $c = 1.45 \text{M}_{\text{L}}$.
\section{Bootstrap data-sets and parameter estimates} \label{app:gpd_parametric_bootstrap}

\subsection{Relation to main text}
This section supports the material presented in Section 2.2 of the main paper. To represent the sampling variability in the maximum likelihood estimate $\bm{\hat \theta}$ associated with the log-likelihood (2) of the main text we take a parametric bootstrapping approach. This appendix describes the simulation of bootstrap catalogues and how they may be used to obtain bootstrap estimates of $\bm{\hat\theta}$. Bootstrap parameter estimates of this type are used throughout the main text. 

\subsection{Generating bootstrapped data-sets}
\subsubsection{Threshold exceedances and a point process}
\begin{figure}
    \centering
    \includegraphics[width = 0.75\textwidth]{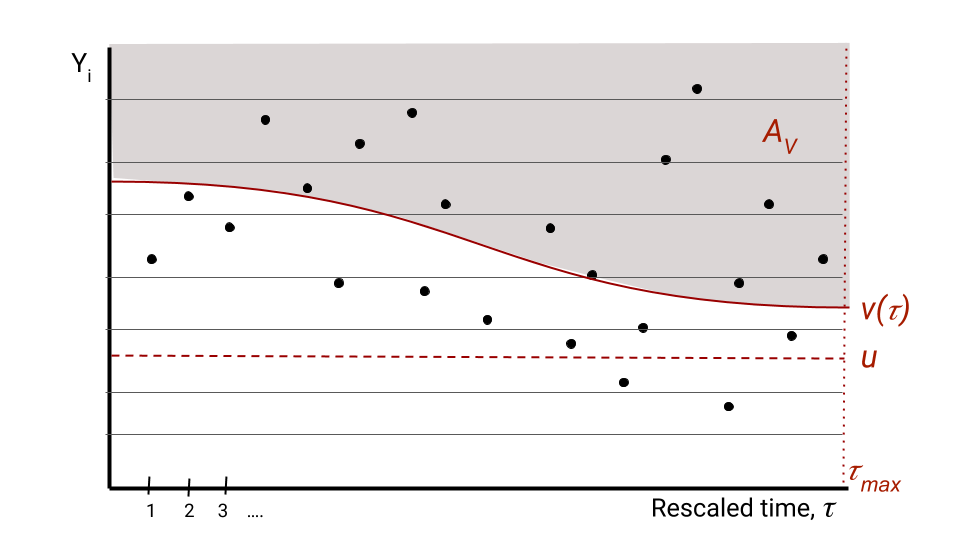}
    \caption{Threshold exceedances as point process}
    \label{fig:bootstrap_point_process}
\end{figure}

In bootstrap realisations of the earthquake catalogue, the number, timing and magnitudes of events exceeding $v(\tau)$ are all variable. In an alternate catalogue, events remain within the transformed observation interval $(0, \tau_{\text{max}})$ but no longer form a regularly spaced sequence. Rather, the events in the region $A_v = \{(\tau, y) : 0 \leq \tau \leq \tau_\text{max} , y \geq v(\tau)\}$, as shown in Figure~\ref{fig:bootstrap_point_process}, are approximated by a Poisson process.

The Poisson process intensity on $A_v$ is determined by the GPD parameters $\bm{\theta}$ and $\lambda_u$, where $\lambda_u$ is the expected number of events exceeding $u \leq \min_{0< \tau < \tau_{\text{max}}} v(\tau)$ per unit $\tau$ \citep{coles2001introduction}. Since it is assumed that no censoring occurs on $A_v$ and that the underlying magnitude distribution is identical over $t$, it follows that $\lambda_u$ is constant over $\tau$ because time has been transformed so that earthquakes occur at a constant rate. The resulting intensity function on $A_v$ is: 
\begin{equation} \label{app:gpd_pp_intensity}
        \lambda(\tau, y) = \frac{\lambda_u}{\sigma_u} \left[ 1 + \xi \frac{y - u}{\sigma_u} \right]^{1/\xi - 1}_+ \quad \text{ for } (\tau, y) \in A_v.  
\end{equation}

Bootstrap catalogues of earthquakes with magnitudes exceeding $v(\tau)$ can be obtained as realisations from the point process with intensity function~\eqref{app:gpd_pp_intensity} using the estimated parameters $\bm{\hat\theta}$. We first describe how to generate such bootstrap catalogues and then how these can be used to obtain bootstrap estimates of $\bm{\hat\theta}$. 

\subsubsection{Simulating the exceedance count}
The first step in generating a bootstrap catalogue is to sample the number of events that occur on $A_v$. The number of events on $A_v$ is Poisson distributed with expectation 
\begin{equation} \label{eqn:pp_integrated_intensity}
     \Lambda(A_v) = \int_{A_v} \lambda(\tau, y) \ \mathrm{d}\tau \ \mathrm{d}y, 
\end{equation}
which must be estimated from the observed catalogue. This is complicated by the rounding of observed magnitudes $\bm{x}$ to the nearest $2\delta$. Recall that for borderline events $\{x_i \in \bm{x}: |x_i - v_i| < \delta\}$ it is not known whether or not the corresponding unrounded magnitudes exceed $v(\tau)$ and place those events on $A_v$. Therefore $n_v$, the observed event count on $A_v$, is unknown and must itself be estimated. 

Given the rounded magnitudes $\bm{x} = (x_1, \dots, x_n)$ and the estimated GPD parameters $(\hat \sigma_u, \hat \xi)$, events $i = 1,\dots, n$ each exceed their magnitude threshold and are on $A_v$ independently with probability $w_i$ as defined in expression (3) of the main text. Uncertainty in the observed event count due to magnitude rounding can be included in the generation of bootstrap sample sizes by simulating a value $m_v$ for $n_v$, as the sum of $n$ independent Bernoulli random variables with expectations $w_1$ to $w_n$. 

This simulated value for the observed event count can be used as a point estimate $\hat\Lambda(A_v) = m_v$ for $\Lambda(A_v)$. Since $\hat\Lambda(A_v)$ is an estimate of based on a single observed count, it is important to also include uncertainty in the inferred value of $\Lambda(A_v)$ when generating bootstrap catalogue sizes. This can be done using a bootstrapped value for the value of the estimator $\hat\Lambda(A_v)$. Since inference is based on a single Poisson count, the bootstrap estimator $\tilde \Lambda(A_v)$ is obtained simply as a sampled value from the Poisson($\hat\Lambda(A_v)$) distribution. 

Finally, the event count $\tilde n$ to be used for the bootstrapped catalogue can be sampled from a Poisson($\tilde\Lambda(A_v)$) distribution. Doing so properly represents rounding, estimation and sampling uncertainties in the bootstrapped event counts. 

\subsubsection{Simulation of event times}
 Having simulated the exceedance count $\tilde n_v$, the times $\bm{\tilde\tau} = (\tau_1, \dots, \tau_{\tilde n_v})$ of the events on $A_v$ can sampled according to the marginal temporal intensity $\lambda(\tau)$. The marginal temporal intensity is found by integrating the joint intensity \eqref{app:gpd_pp_intensity} over magnitudes. Noticing that $\lambda(\tau, y)$ is proportional to the GPD density, $\lambda(\tau)$ can be stated in terms of the GPD survivor function 
 $\bar F(y; \sigma, \xi) = [1 + \xi y /\sigma]_+^{-1/\xi}$ to give:
\begin{equation} \label{eqn:pp_temporal_intensity}
    \lambda(\tau) 
    = \int_{v(\tau)}^{\infty} \lambda(\tau, y) \ \mathrm{d}y 
    = \lambda_u \bar F(v(\tau) - u; \sigma_u , \xi) \quad \text{ for } 0 \leq \tau \leq \tau_{\text{max}}.
\end{equation}
Sampling exceedance times from this intensity can be achieved through reverse application of the time-rescaling theorem \citep{brown2002time}. In general, this will require numerical integration and can be computationally intensive. Depending on the form of $v(\tau)$ more efficient methods may be available. When $v(\tau)$ is a step function the step-wise integral of the temporal intensity has a simple form and event times can be simulated very efficiently; the $\tilde n_v$ events are allocated independently to steps with probability proportional to the temporal intensity integrated over each step. Events are then be located uniformly at random within their allocated step. When a more complex form is used for $v(\tau)$, alternative sampling approaches may reduce the computational cost of sampling event times. For example, exact samples may be obtained through rejection sampling and approximate samples by approximating $v(\tau)$ as a step function. 

\subsubsection{Conditional simulation of event magnitudes}
The magnitude of each event in the bootstrap catalogue may then be simulated conditional on its occurrence time. Given a simulated occurrence time $\tilde \tau \in (0, \tau_{\text{max}})$, the conditional magnitude intensity for magnitudes exceeding $v(\tilde \tau)$ is simply the GPD density function with shape parameter $\xi$ and time-dependent scale parameter $\sigma(\tilde \tau) = \sigma_u + \xi(v(\tilde \tau) - u)$: 
\begin{equation} \label{eqn:pp_conditional_magnitude_intensity}
    \lambda(y|\tilde \tau) = \frac{\lambda(\tilde \tau, y)}{  \lambda(\tilde \tau)} = \frac{1}{\sigma(\tilde \tau)} \left[ 1 + \xi \frac{y - v(\tilde \tau)}{\sigma(\tilde \tau)}\right]_{+}^{-1/\xi -1}, \quad \text{ for } y \geq v(\tilde \tau). 
\end{equation}
This allows easy simulation of magnitudes $\bm{\tilde y} = (\tilde y_1, \dots, \tilde y_{\tilde n_v})$ conditional on their occurrence times $\bm{\tilde \tau}$, by generating random variates from the appropriate GPD. The simulation of a bootstrap earthquake catalogue is completed by rounding these to the nearest multiple of $2\delta$, to obtain $\bm{\tilde x} = (\tilde x_1, \dots, \tilde x_{\tilde n_v})$. Note that in a bootstrap catalogue $\tilde y_i > \tilde v_i$ for $i = 1,\dots, \tilde n_v$, but following rounding it is possible that some of the $\tilde x_i$ are below their associated threshold value. The simulation of bootstrapped datasets is summarised in Algorithm~\ref{alg:simulate_bootstrap_catalogues}, where $\hat \sigma(\tilde \tau) = \hat\sigma_u + \hat\xi(v(\tilde \tau) - u)$.

\begin{algorithm}[htbp]
\SetAlgoLined
\KwResult{A bootstrapped dataset of rounded GPD observations $\bm{\tilde x}$, based on the threshold function $v(\tau)$, observations $\bm{x}$ and parameter estimates $\hat{\bm{\theta}}$.}
 Sample $m_v$ as the sum of independent Bernoulli($w_i$) realisations where $i = 1,\dots,n$ \;
 Sample the estimate of $\Lambda(A_v)$, $\tilde \Lambda(A_v)$ from the Poisson($m_v$) distribution\;
 Sample the bootstrap number of exceedances $\tilde n_v$ from the Poisson($\tilde\Lambda(A_v)$) distribution\;
 \For{$i=1$ \KwTo $i =\tilde n_v$}{
    sample $u_i$ from a Uniform(0,1) distribution\; 
    find the bootstrap occurrence time $\tilde \tau_i$ which satisfies 
        \begin{equation*} \label{eqn:app_sample_times_step}
        \int_0^{\tilde\tau_i} \lambda(\tau; \hat \sigma_u, \hat\xi, v(\tau)) \ \mathrm{d} \tau  = u_i \hat \Lambda(A_v);
        \end{equation*} \\
    sample the bootstrap magnitude exceedance $\tilde z_i$ from the $\text{GPD}(\hat\sigma(\tilde \tau_i), \hat\xi)$ distribution\;
    calculate the bootstrap latent magnitude $\tilde y_i$ = $v(\tilde \tau_i) + \tilde z_i$\;
    round $\tilde y_i$ to the nearest $2\delta$ to get the rounded bootstrap magnitude $\tilde x_i$ (which may be less than $v(\tilde\tau_i)$).
    }
 \caption{Simulation of GPD data with variable threshold and rounding}
 \label{alg:simulate_bootstrap_catalogues}
\end{algorithm}

\subsection{Generating bootstrapped maximum likelihood estimates} 
In the bootstrapped earthquake catalogues, some magnitudes $\tilde{\bm{x}}$ may be less than their respective threshold values. However, unlike in the original catalogue, each of the corresponding unrounded magnitudes $\tilde{\bm{y}}$ exceeds the respective modelling threshold. Therefore, an unweighted log-likelihood should be used when obtaining bootstrap maximum likelihood estimates. Letting $\tilde v_i = v(\tilde \tau_i)$ and $\sigma_{\tilde v_i} = \sigma_u + \xi(\tilde v_i - u)$, the unweighted log-likelihood function is
 \begin{equation*} \label{eqn:app_unwieghted_rounded_gpd_llh}
    \ell(\bm{\theta} | \bm{\tilde x}, \bm{\tilde v})
     = \sum_{i = 1}^{\tilde n_v} 
    \log\left[ 
         F(\tilde x_i + \delta - \tilde v_i; \sigma_{\tilde v_i}, \xi) - F(\max(\tilde v_i, \tilde x_i - \delta) - \tilde v_i; \sigma_{\tilde v_i}, \xi)
        \right]. 
 \end{equation*}
The maximum likelihood estimates resulting from a collection of bootstrap catalogues can be used to represent the sampling uncertainty of the original maximum likelihood point estimate $\hat{\bm{\theta}}$. This is done in the main text when calculating confidence intervals on parameter values, conditional quantiles and return levels. The bootstrap parameter estimates are also used in the construction of adapted PP and QQ plots and when evaluating metric values. 

\section{Sampling standardised threshold exceedances} \label{app:sampling_z}

\subsection{Connection to main text}
 This appendix describes how to sample a vector $\tilde{\bm{z}}$ of unrounded threshold exceedances transformed to have a common Exp(1) marginal distribution. This used in Sections 3-6 of the main text and requires: a single bootstrap estimate $\tilde{\bm{ \theta}}$ of the estimated GPD parameters $\hat{\bm{\theta}}$ (obtained as described in Appendix~\ref{app:gpd_parametric_bootstrap}), the $n$-vector of rounded observations $\bm{x}$ and the corresponding threshold vector $\bm{v}$. The process for sampling a single vector $\tilde{\bm{z}}$ is described and is formalised in Algorithm~\ref{alg:simulate_standardised_exceedances}. 
 
 \subsection{Sampling unrounded threshold exceedances}
 It is unknown which, if any, of the borderline values $\{x_i \in \bm{x}: |x_i - v_i| < \delta\}$ correspond to unrounded values $y_i \in \bm{y}$ that exceed the modelling threshold $v(\tau)$. The first step in sampling $\tilde{\bm{z}}$ is therefore to sample the set $I$ that exceed the modelling threshold. This is done by simulating independent Bernoulli trials for each event $i = 1, \dots, n$ with success probabilities $w_i = \Pr(Y_i > v_i | x_i, \hat{\bm{\theta}})$ as defined in equation (3) of the main text.  The vector $\tilde{\bm{z}}$ will therefore have a randomly sampled length $\tilde m = |I| \leq n$, where the distribution of $\tilde m$ depends on $\bm{x}, \bm{v}$ and $\tilde{\bm{\theta}}$. 
 
 The unrounded magnitude values for events in $I$ are then simulated from their conditional distribution given: their rounded values, the estimated GPD parameters and that they are threshold exceedances. Letting $F$ be the GPD distribution function as in Equation (1) of the main text, the required conditional distribution function of $Y_i| x_i, \bm{\theta}, Y_i \geq v_i$ is given by: 
\begin{equation} \label{eqn:unrounded_conditional_distribution}
    G_{Y_i| x_i, \bm{\theta}, Y_i \geq v_i}(y) = 
    \left\{
    \begin{array}{cr}
        0 & \text{ for } y < b_i,  \\
        \frac{F(y - u;\bm{\theta}) - F(b_i - u; \bm{\theta})}{F(x_i + \delta - u; \bm{\theta}) - F(b_i - u ; \bm{\theta})} & \text{ for } b_i \leq y \leq x_i + \delta, \\
        1 & \text{ for } y > x_i + \delta,
    \end{array}
    \right.
\end{equation}
where $i \in I$ and $b_i = \max(x_i - \delta, v_i)$ is the smallest value above the modelling threshold that results in the rounded observation $x_i$. The sampled values are combined to create $\tilde{\bm{y}}$, an $\tilde{m}$-vector of continuous-valued sampled threshold exceedances. Note that the length of this vector is a random variate when there are one or more borderline values. By repeated simulation using $k$ bootstrap parameter estimates $\{\tilde{\bm{\theta}}^{(1)}, \dots, \tilde{\bm{\theta}}^{(k)}\}$, the resulting vectors of unrounded exceedances $\{\tilde{\bm{y}}^{(1)}, \dots, \tilde{\bm{y}}^{(k)}\}$ reflect uncertainty about both the GPD parameter values and the number of threshold exceedances. 

\subsection{transformation onto common margins}
When $v(\tau)$ is non-constant, the elements of $\tilde{\bm{y}}$ are random variates from the GPD family, but they do not share a common set of parameters. To resolve this issue, the probability integral transform can then be used to give each element of $\tilde{\bm{y}}$ an Exp(1) marginal distribution using its fitted GPD parameters. This results in a vector $\tilde{\bm{z}}$ of standardised, sampled exceedances of the modelling threshold $v(\tau)$. The bootstrap simulation of one such vector is formalised in Algorithm~\ref{alg:simulate_standardised_exceedances}.

Note that the transformation onto exponential margins is dependent on the estimated GPD parameters. The effect of parameter uncertainty on this transformation is represented across bootstrap catalogues by transforming each collection of bootstrapped threshold exceedances $\{ \tilde{\bm{y}}^{(1)}, \dots, \tilde{\bm{y}}^{(k)} \}$ using their respective bootstrapped parameter estimates $\{ \tilde{\bm{\theta}}^{(1)}, \dots, \tilde{\bm{\theta}}^{(k)} \}$. This yields a set of $k$ sampled vectors of threshold exceedances on exponential margins, $\{ \tilde{\bm{z}}^{(1)}, \dots, \tilde{\bm{z}}^{(k)} \}$. As with $\tilde{\bm{y}}$, the length of $\tilde{\bm{z}}$ is a an $\tilde m$-vector and so is a random variate when there are one or more borderline events in $\bm{x}$. 

These sampled vectors of standardised threshold excesses can be used to calculate expected metric values or to construct modified PP- and QQ-plots, as in Figures (3) and (8) of the main text. In those plots, the variability between the empirical quantiles (or probabilities) of each $\{ \tilde{\bm{z}}^{(1)}, \dots, \tilde{\bm{z}}^{(k)} \}$ is shown by the confidence intervals on sample quantile values (or probabilities). The expected range of values for each sample quantile (or probability) is shown by the tolerance intervals. The uncertainty in the number of threshold exceedances is also incorporated when calculating the tolerance intervals; they are constructed using $k$ sets of Exp(1) random variates of lengths $\dim \tilde{\bm{z}}^{(1)}, \dots, \dim \tilde{\bm{z}}^{(k)}$.

\begin{algorithm}[htbp]
\SetAlgoLined
    \SetKwInOut{Input}{input}
    \SetKwInOut{Output}{output}
    \Input{A bootstrap estimate $\tilde{\bm{\theta}} = (\tilde\sigma, \tilde\xi)$ of the GPD parameters, the $n$-vector of rounded observed values $\bm{x}$, and their corresponding thresholds $\bm{v}$.}
    \Output{A vector $\tilde{\bm{z}}$ of length $\tilde m \leq n$ of sampled unrounded values, transformed to have an Exp(1) distribution under the fitted model.}
  
    \For{$i = 1$ \KwTo $n$}{
    calculate $w_i = \text{Pr}(Y_i > v_i | x_i, \tilde{\bm{\theta}})$, the probability that each rounded observation corresponds to an unrounded value on $A_v$, as in Equation (3) of the main text\;
    }
    Generate $n$ independent Uniform[0,1] random variates $u_1,\dots,u_n$ \;
    Sample the indexing set of events that are on $A_v$, $I = \{i \in (1,\dots,n) : u_i \leq w_i\}$ and let $\tilde m = |I|$ \;
    Store the elements of $I$ in the vector $\bm{\beta}= (\beta_1, \dots, \beta_{\tilde m})$ so that $\tilde y_{\beta_i} > v_{\beta_i}$ for $i = 1,\dots, \tilde m$ and initialise $\bm{\tilde y}$ and $\bm{\tilde z}$ as vectors of length $\tilde{m}$\;
    \For{$j = 1$ \KwTo $\tilde m$}{
        Let $a = \beta_{j}$ \;
        Sample the $j^{\text{th}}$ unrounded exceedance $\tilde{y}_{j}$ from its conditional distribution $G_{Y_a | x = x_a,\bm{\theta} = \tilde{\bm{\theta}},  Y_a \geq v_a,} (y)$ as in equation \eqref{eqn:unrounded_conditional_distribution} \;
        Let $\tilde{\bm{\theta}}_{v_a} = (\tilde \sigma - \tilde\xi(v_a - u), \tilde\xi)$ be the bootstrapped GPD parameters for exceedances of $v_a$\;
        Transform $\tilde{y}_j$ onto  Exp(1) margins under this fitted model by letting $F$ be the GPD distribution function (1) in the main text and setting 
        $$\tilde{z}_j = - \log \left[ 1 - F\left(y_j - v_{a}; \tilde{\bm{\theta}}_{v_a} \right)\right].$$
    }

 \caption{Simulation of standardised threshold exceedance sets}
 \label{alg:simulate_standardised_exceedances}
\end{algorithm} 

\section{Code}
An R package is currently under development to implement the methods introduced in this paper. Unpolished R code used for the analyses of the main text can be found at \url{https://github.com/zakvarty/threshold_paper_code}.

\section{Supplementary figures}


\begin{figure}
    \centering
        \includegraphics[width = 0.45\textwidth, page = 1]{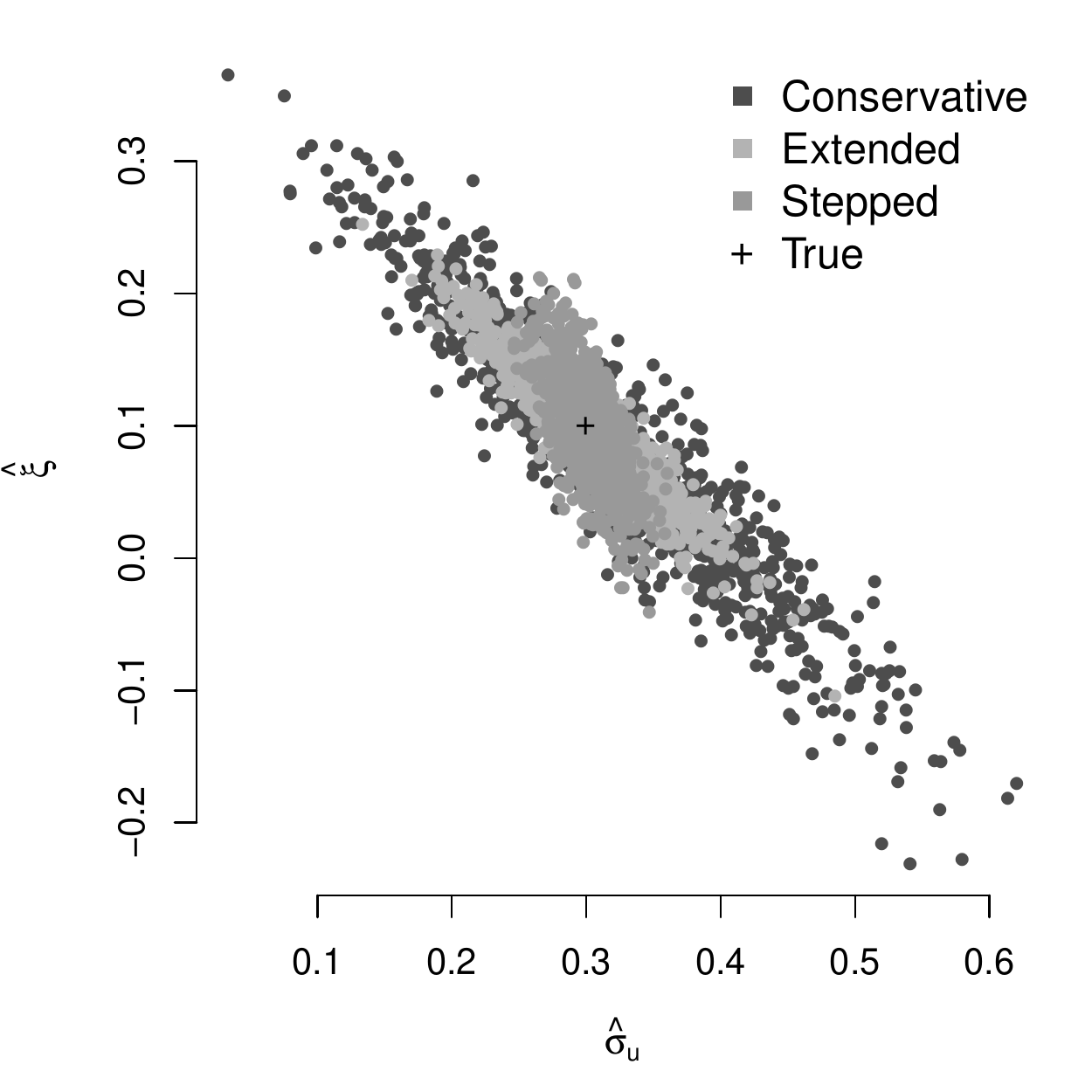}
         \includegraphics[width = 0.45\textwidth, page = 1]{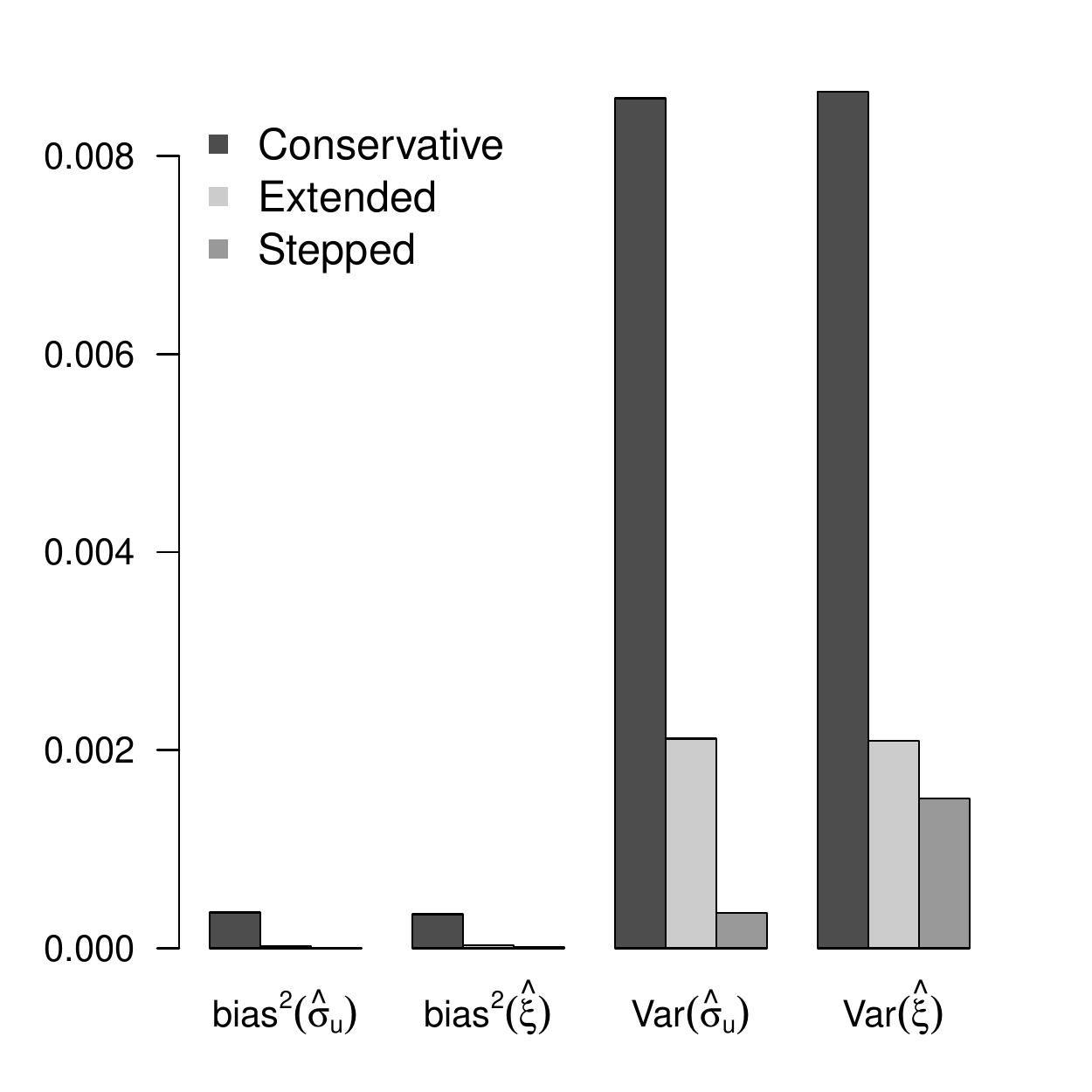}
    \caption{[Left] Plot of maximum likelihood estimates of GPD scale and shape parameters for 1000 simulated catalogues obtained using a conservative, stepped and extended approach. [Right] Plot of mean squared error decomposition for maximum likelihood estimates by each approach. The error is decomposed into squared bias and variance terms for each of the GPD parameters, with variance terms having larger contributions.}
    \label{fig:sim_cat_mse}
\end{figure}

\begin{figure}
    \centering
    \includegraphics[width = 0.45 \textwidth, page = 3]{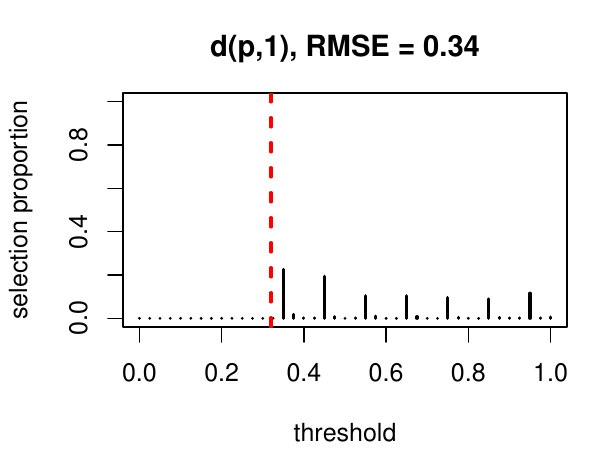}
    \includegraphics[width = 0.45\textwidth, page = 4]{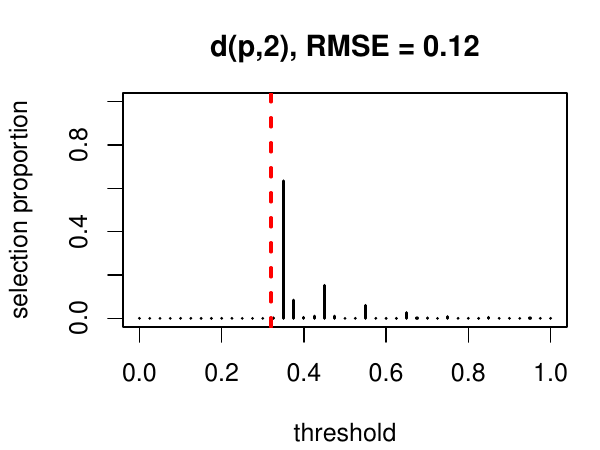}
    \caption{Sampling distribution of threshold selection methods for PP-based metrics over 500 simulated catalogues with constant threshold and hard censoring. The true threshold is shown by a dashed red line and the root mean squared error (RMSE) for each method is given in plot titles.}
    \label{fig:flat_threshold_selection_pp_summary}
\end{figure}

\begin{figure}
    \centering
    \includegraphics[width = 0.45\textwidth]{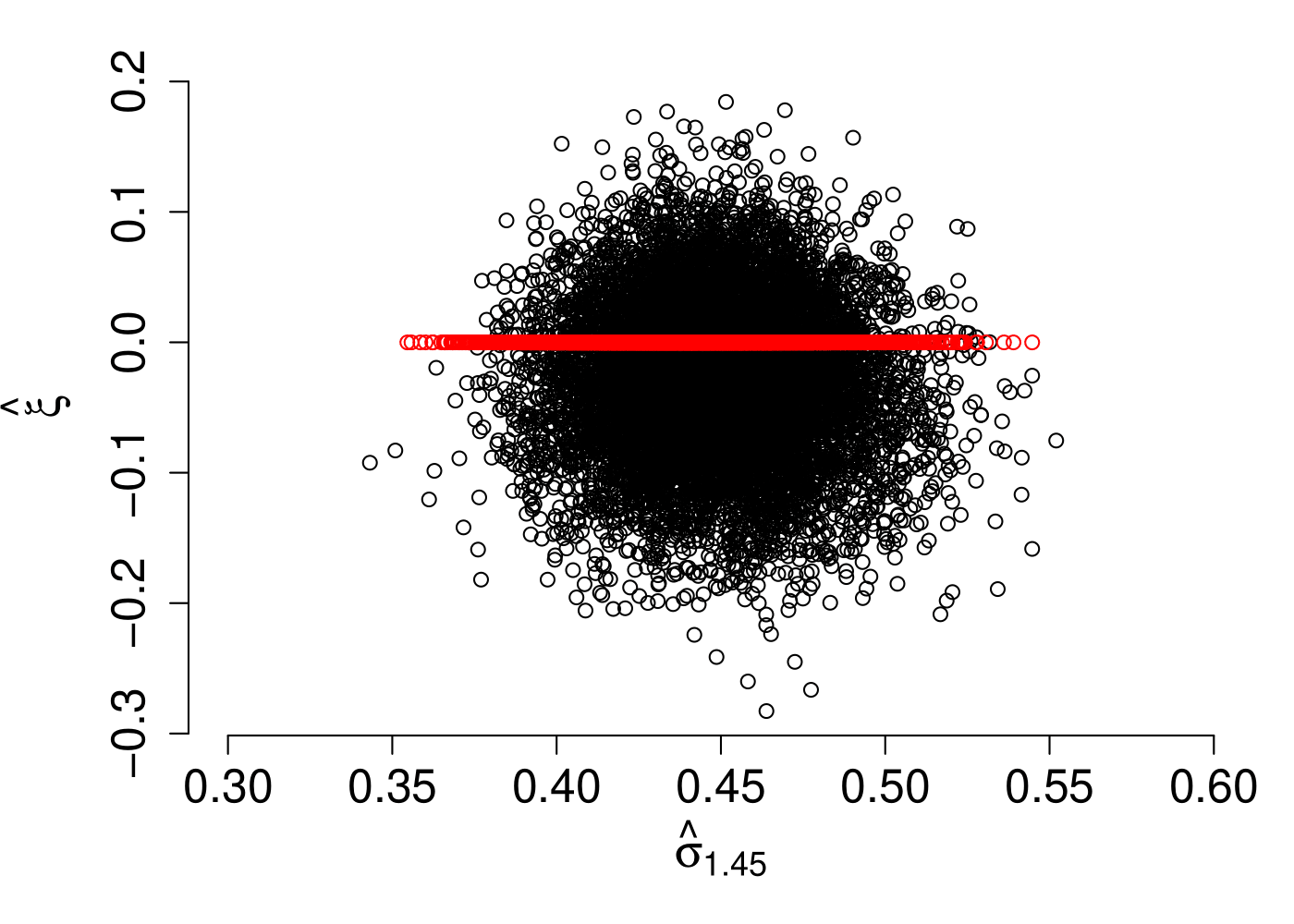}
    \includegraphics[width = 0.45\textwidth]{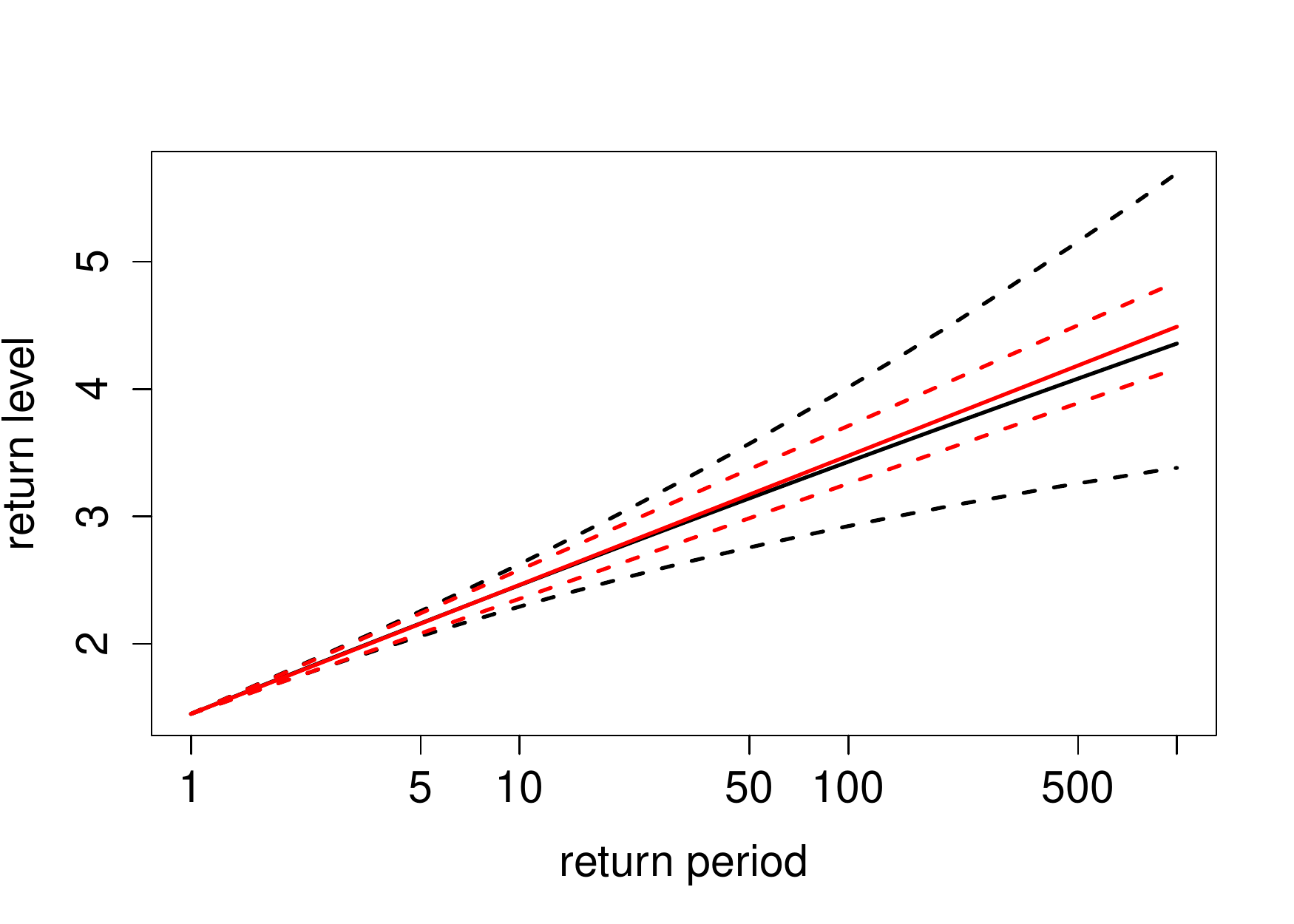}
    \caption{[left] Bootstrap maximum likelihood estimates for GPD parameters for Groningen magnitudes above 1.45ML assuming GPD (black) and exponential (red) models. [right] Resulting conditional return level plot with return period measured in number of events exceeding 1.45 ML.}
    \label{fig:groningen_mles_and_return_levels_above_145}
\end{figure} 

\vspace{-5cm}
\begin{figure}
    \centering
    \includegraphics[width = 0.8\textwidth]{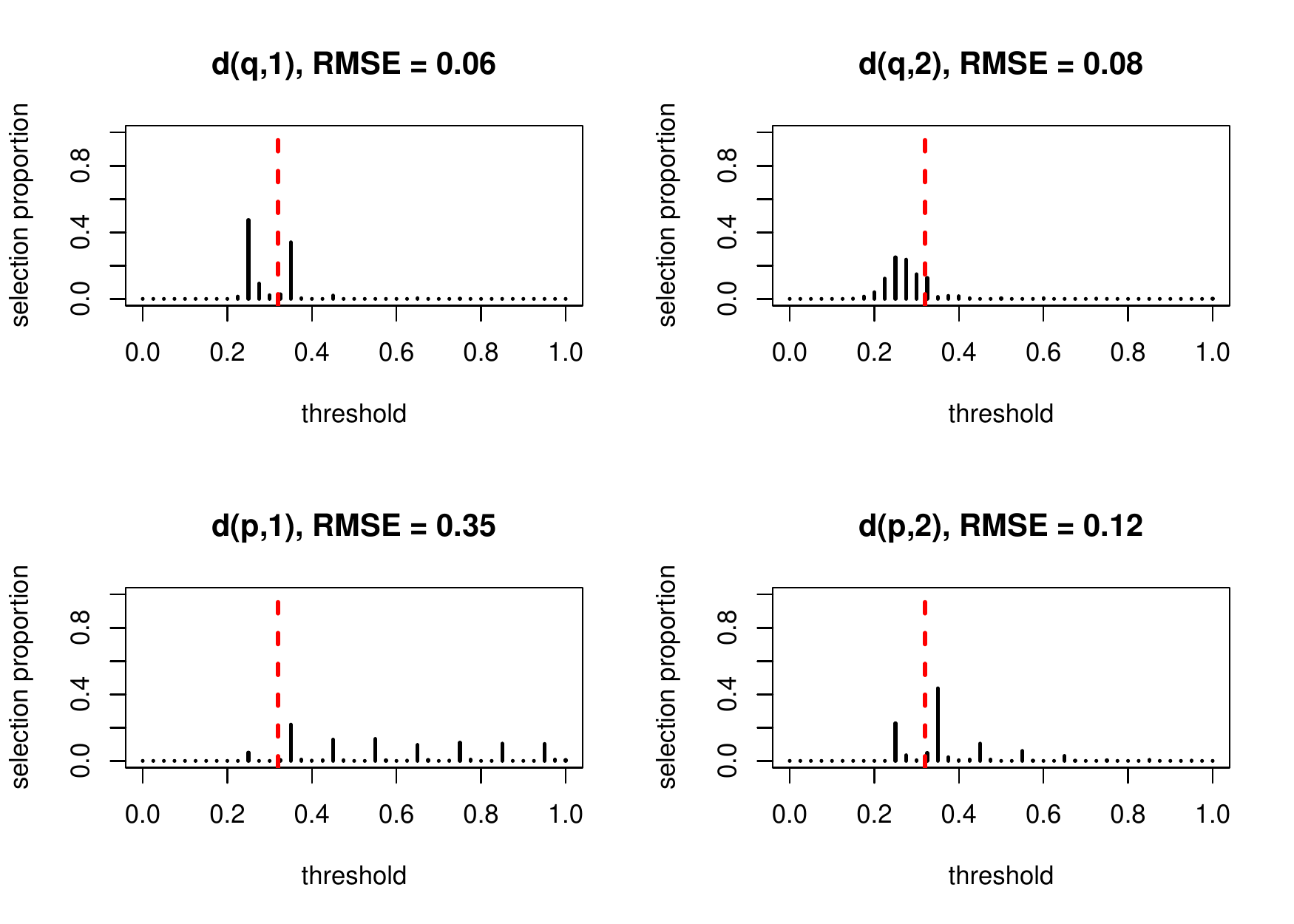}
    \caption{Sampling distribution of threshold selection methods for QQ-based (top row) and PP-based (bottom row) metrics over 500 simulated catalogues with constant threshold and phased censoring. The true threshold is shown by a dashed red line and the root mean squared error (RMSE) for each method is given in plot titles.}
    \label{fig:flat_threshold_selection_pp_qq_summary_phased}
\end{figure}

\FloatBarrier
\end{appendix}

\bibliographystyle{imsart-nameyear} 
\bibliography{sm_references}       
